\documentclass[12pt]{iopart}
\usepackage{iopams}  
\usepackage{setstack}

\begin{document}
\title{On a method for constructing the Lax pairs for nonlinear integrable equations}

\author{I T Habibullin$^{1,2}$, A R Khakimova$^{2}$ and M N Poptsova$^1$}

\address{$^1$Ufa Institute of Mathematics, Russian Academy of Science,
112, Chernyshevsky Street, Ufa 450008, Russian Federation}
\address{$^2$Bashkir State University, 32 Validy Street, Ufa 450076 , Russian Federation} 
\eads{\mailto{habibullinismagil@gmail.com}, \mailto{aigulya.khakimova@mail.ru} and \mailto{mnpoptsova@gmail.com}}

\begin{abstract}

We suggest a direct algorithm for searching the Lax pairs for nonlinear
integrable equations. It is effective for both continuous and discrete models.
The first operator of the Lax pair corresponding to a given nonlinear equation
is found immediately, coinciding with the linearization of the considered
nonlinear equation. The second one is obtained as an invariant manifold to the
linearized equation. A surprisingly simple relation between the second
operator of the Lax pair and the recursion operator is discussed: the recursion
operator can immediately be found from the Lax pair. Examples considered in
the article are convincing evidence that the found Lax pairs differ from the
classical ones. The examples also show that the suggested objects are true Lax
pairs which allow the construction of infinite series of conservation laws and
hierarchies of higher symmetries. In the case of the hyperbolic type partial
differential equation our algorithm is slightly modified; in order to construct
the Lax pairs from the invariant manifolds we use the cutting off conditions for
the corresponding infinite Laplace sequence. The efficiency of the method is
illustrated by application to some equations given in the Svinolupov-Sokolov
classification list for which the Lax pairs and the recursion operators have not
been found earlier.

\end{abstract}


\maketitle

\eqnobysec

\section{Introduction}

In the present article we suggest an algorithm for constructing the Lax pairs for nonlinear integrable models. Our scheme is based on the symmetry approach. Let us explain the algorithm with the evolutionary type PDE of the form 
\begin{eqnarray} 
u_t =f(x,t,u,u_1,u_2,...,u_k), \qquad u_j=\frac{\partial^{j} u}{\partial x^j}. \label{I1}
\end{eqnarray}
Recall that equation
\begin{eqnarray} 
u_{\tau} =g(x,t,u,u_1,u_2,...,u_m)   \label{I2}
\end{eqnarray}
is called a symmetry for the equation (\ref{I1}) if function $ g $ satisfies the following differential equation 
\begin{eqnarray} 
 \left(D_t - \frac{\partial f}{\partial u} - \frac{\partial f}{\partial u_1} D_x - ... -\frac{\partial f}{\partial u_k} D_x^k\right) g = 0  \label{I3}
\end{eqnarray}
where $D_t$ and $D_x$ are the operators of the total derivatives with respect to $t$ and $x$ correspondingly. Note that equation (\ref{I3}) is overdetermined and for any fixed value of $ m $ one can effectively find all of its solutions of the form $g=g(x,t,u,u_1,u_2,...,u_m)$ (see \cite{IbragimovShabat}).

An ordinary differential equation 
\begin{eqnarray} 
u_m =G(x,t,u,u_1,u_2,...,u_{m-1})   \label{I4}
\end{eqnarray}
defines an invariant manifold for (\ref{I1}) if the following condition is satisfied 
\begin{eqnarray} 
D_t G - D^m_x f|_{(1.1),\,(1.4)} = 0. \label{I5}
\end{eqnarray}
Here $ D_t G $ is evaluated by means of the equation (\ref{I1}) and all the $x$-derivatives of the order greater than $m-1$ are expressed due to the equation (\ref{I4}) and its differential consequences. Equation (\ref{I5}) generates a PDE with unknown $ G $ admitting a large class of solutions. However it is a hard problem to find these solutions since the equation is not overdetermined. Some of the invariant manifolds to (\ref{I1}) can be found from the stationary part of the symmetry (\ref{I2}) by taking $ g(x,t,u,u_1,...,u_m) = 0. $
 
We now concentrate on the linearization of the equation (\ref{I1}) around its arbitrary solution $ u = u(x,t): $ 
\begin{eqnarray} 
v_t - \frac{\partial f}{\partial u} v - \frac{\partial f}{\partial u_1} v_1 - \dots-\frac{\partial f}{\partial u_k} v_k = 0 . \label{I6}
\end{eqnarray} 
Actually (\ref{I6}) defines a family of differential equations, depending on $u$. The important fact that $u$ ranges the whole set of the solutions to (\ref{I1}) is formalized in the following way. We assume that in addition to the natural independent dynamical variables $ v, v_1, v_2, ... $ the variables $u, u_1, u_2, ... $ are also considered as independent ones.

Let us find the invariant manifold of the form 
\begin{eqnarray}
v_m = \sum^{m-1}_{j=0} a(j) v_j , \quad v= v_0, \label{I7}
\end{eqnarray}
to the equation (\ref{I6}). This means that the condition
\begin{eqnarray} 
D_t \left(\sum^{m-1}_{j=0} a(j) v_j\right) - D^m_x \left(\sum^{k}_{j=0} \frac{\partial f}{\partial u} v_j\right)|_{(1.1),\, (1.6),\, (1.7)} = 0  \label{I8}
\end{eqnarray}
holds identically for all values of $ u, u_1, u_2, \dots . $
Here we assume that $ \forall  j$ function $a(j) $ depends on $ x, t $ and a finite number of the dynamical variables $ u, u_1, u_2, ... $. 
In (\ref{I8}) the variables $ u_t, u_{xt}, u_{xxt}, ...  $ are expressed by means of the equation (\ref{I1}) and the variables $v_t$, $v_{1,t}$ ,..., $v_{m-1,t}$ are exresssed due to (\ref{I6}). Equation (\ref{I8}) splits down into a system of $m$ partial differential equations with the unknown functions $a(0),a(1),\dots, a(m-1)$, coefficients of the decomposition (\ref{I7}). Due to the presence of the additional dynamical variables $u,u_1,u_2,...$ the system is overdetermined and hence can effectively be investigated. Actually, equation (\ref{I7}) defines a bundle of the manifolds depending on infinite set of the variables $u, u_1, u_2, ... $.
Suppose that such an invariant manifold is found. Then we can interpret a pair of the equations (\ref{I6}), (\ref{I7}) as the Lax pair to the equation (\ref{I1}). In the examples the order $k$ of the equation (\ref{I1}) and the order $m$ of the equation (\ref{I7}) coincide.

As an illustrative example we consider the well known KdV equation
\begin{equation}
\label{s1kdv}
u_{t}=u_3+uu_1.
\end{equation}
Linearized equation 
\begin{equation} \label{s1KdV:lin}
v_t = v_{3} + u v_1 + u_1 v
\end{equation}
obtained by virtue of the rule (\ref{I6})
admits the third order invariant manifold defined by the equation
\begin{equation}\label{s1KdV:L}
v_{3} = \frac{u_{2}}{u_1} v_{2} - \left( \frac{2}{3} u + \lambda\right) v_1 + \left( \left( \frac{2}{3} u + \lambda \right)\frac{u_{2}}{u_1} - u_1 \right)v.
\end{equation}
It is easily checked that equations (\ref{s1KdV:lin}) and (\ref{s1KdV:L}) are consistent if and only if the function $u=u(x,t)$ satisfies the equation (\ref{s1kdv}). Therefore these equations constitute the Lax pair to the KdV equation. It  differs from the usual one found earlier in \cite{Lax68}. Stress that there is no any second order invariant manifold of the form $v_{2}=a(u,u_1,u_2)v_{1}+b(u,u_1,u_2)v$ for the equation (\ref{s1KdV:lin}). But there is a first order one $v_1=\frac{u_2}{u_1}v$ which however does  not contain $\lambda$ and therefore does not generate any true Lax pair. 

Recursion operator is an important attribute of the integrability theory. It gives a compact description for the hierarchies of both symmetries and conservation laws. Various methods for studying the recursion operators can be found in literature (see, for instance \cite{GursesKarasu}-\cite{mikhailov} and the references therein).

We observe that invariant manifold for the linearized equation is closely connected with the recursion operator for the original equation. Indeed examples in \S6 show that equation defining the invariant manifold, which provides the second operator of the Lax pair, can be rewritten as a formal eigenvalue problem of the form
$$Rv=\lambda v$$
for the recursion operator $R$.
For instance, equation (\ref{s1KdV:L}) is easily rewritten as (see \S 6)
$$(D_x^2+\frac{2}{3}u+\frac{1}{3}u_xD_x^{-1})v=\lambda v$$
where the operator at the l.h.s. is nothing else but the recursion operator for the KdV hierarchy.

Therefore our scheme of constructing the Lax pairs provides an alternative tool for searching the  recursion operator. On the other hand side when the recursion operator is known the invariant manifold and hence the Lax pair can be found by simple manipulations (see \S 6). At this point we have an intersection with the pioneering articles \cite{IbragimovShabat}, where nonstandard Lax pairs are given for some of the KdV type equations in terms of the recursion operators. However these Lax pairs differ from those found within our scheme since they are nonlocal and do not contain any spectral parameter.

There is a great variety of approaches for searching the Lax pairs from the Zakharov-Shabat dressing \cite{ZakharovShabat74,ZakharovShabat79} and prolongation structures by Wahlquist and Estabrook \cite{WahlquistEstabrook}  to 3D consistency approach developed in \cite{NijhoffWalker}-\cite{Nijhoff}. We mention also approaches proposed in  \cite{Yamilov82, Xenitidis, IbragimovShabat}. An advantage of our scheme is that it can be applied to any integrable model (at least in 1+1 -dimensional case), the first of the operators is easily found and the second is effectively computed. They allow finding the conservation laws, higher symmetries and invariant surfaces for the corresponding nonlinear equation. The found Lax pairs   are more complicated since they are based on the differential operators of the orders greater than usual ones. This is their disadvantage. The question remains open whether the Lax pairs of this kind allow to find any new solution for the well studied models.

In the case of the hyperbolic type integrable equations the algorithm should be slightly modified. Let us explain it with the example of the sine-Gordon equation
\begin{equation} 
\label{s1.sin}
u_{xy}=\sin u.
\end{equation}
Let us find the simplest but nontrivial, i.e. depending on a parameter, invariant manifold for the linearized equation 
\begin{equation} 
\label{s1.lin-sin}
v_{xy}=(\cos u) v.
\end{equation}
It is the three dimensional surface in the space of the dynamical variables $v,v_x,v_y,v_{xx},v_{yy},...$ defined by the following two linear equations, with the coefficients, depending on the field variable $u(x,t)$ and its derivatives 
\begin{equation} \label{s1.sin:inv1}
v_{yy}  - u_y(\cot u) v_y + \lambda \frac{u_y}{\sin u} v_x - \lambda v = 0,
\end{equation}
\begin{equation} \label{s1.sin:inv2}
v_{xx}  - u_x(\cot u) v_x + \lambda^{-1} \frac{u_x}{\sin u} v_y - \lambda^{-1} v = 0.
\end{equation}
Here $\lambda$ is a complex parameter. Note that equations (\ref{s1.sin:inv1}), (\ref{s1.sin:inv2}) are not independent. One of them is immediately found from the other by differentiation by means of the equations (\ref{s1.sin}), (\ref{s1.lin-sin}).  A triple of the equations (\ref{s1.lin-sin})-(\ref{s1.sin:inv2}) can be rewritten (see the end of \S 3) as a pair of the systems of ordinary differential equations providing the Lax pair realized in $3\times3$ matrices. The method for deriving the Lax pair from the triple (\ref{s1.lin-sin})-(\ref{s1.sin:inv2}) is based on constructing the infinite Laplace cascade for the linearized equation (\ref{s1.lin-sin}) and obtaining the finite reduction of the cascade.

Let us give a brief comment on the structure of the article. In \S\S 2,3 we discuss the well-known Laplace cascade for linear and nonlinear hyperbolic type equations. In \S 3 also the problem of finding finite reductions of the infinite Laplace sequence is studied. The Lax pair to the sine-Gordon equation is derived from the Laplace cascade. In \S 4 the definition of the invariant manifold for the hyperbolic type equations is recalled. The Lax pair is construted via invariant manifolds for  hyperbolic equation (\ref{eq12:main}) found in \cite{SokolovMeshkov}.
In \S 5 the Lax pairs are constructed by evaluating invariant manifolds for the evolutionary type integrable equations. Beside the explanatory examples here we consider two equations (\ref{eq12:sym1}) and (\ref{eq12:sym2}) found in \cite{SvinolupovSokolov} as equations possessing infinite hierarchies of conserved quantities. To the best of our knowledge the Lax pairs for the equations (\ref{eq12:main}), (\ref{eq12:sym1}) and (\ref{eq12:sym2}) have never been found before. In \S6 we illustrate applications of the newly found Lax pairs. The Lax pair obtained in the previous sections is used to construct conservation laws for a Volterra type chain. We also show that the second operators of our Lax pairs are closely connected with the recursion operators for the associated nonlinear equations.   In Appendix we give all of the computational details appeared when we evaluated the invariant manifold for the linearization of the sine-Gordon equation.

\section{Laplace cascade for the linear hyperbolic type equations}

Let us recall the main steps of the Laplace cascade method (see \cite{Darboux}, \cite{Goursat}). Consider a linear second order hyperbolic type PDE of the form 
\begin{equation}
\label{s2eq1}
v_{xy}+a(x,y)v_x+b(x,y)v_y+c(x,y)v=0.
\end{equation}
It can easily be checked that functions 

\begin{equation}
\label{s2eq2}
h_{[0]}=a_x+ab-c,\quad k_{[0]}=b_y+ab-c
\end{equation}
do not change under the linear transformation $ v\rightarrow \lambda(x,y)v $ with arbitrary smooth factors $ \lambda(x,y) $ applied to equation (\ref{s2eq1}). They are called the Laplace invariants for (\ref{s2eq1}).

We rewrite equation (\ref{s2eq1}) as a system of two equations: 
\begin{equation}
\label{s2eq3}
\left( \frac{\partial}{\partial y}+a \right)v=v_{[1]}, \quad \left(\frac{\partial}{\partial x}+b\right)v_{[1]}=h_{[0]}v. 
\end{equation}

When the invariant $ h_{[0]} $ does not vanish then one can exclude $ v $ from (\ref{s2eq3}) and obtain a linear PDE for $ v_{[1]} $
\begin{equation}
\label{s2eq4}
v_{[1]xy}+a_{[1]}v_{[1]x}+b_{[1]}v_{[1]y}+c_{[1]}v_{[1]}=0 
\end{equation}
where the coefficients are evaluated as follows 
\begin{equation} 
\label{s2eq5}
a_{[1]}=a-\frac{\partial}{\partial y}\log(h_{[0]}),\quad b_{[1]}=b,\quad c_{[1]}=a_{[1]}b_{[1]}+b_{[1]y}-h_{[0]}.
\end{equation}
Thus we define a transformation of the equation (\ref{s2eq1}) into the equation (\ref{s2eq4}). This transformation is called the Laplace $y$-transformation. Iterations of the transformation generate a sequence of the equations 
\begin{equation}
\label{s2eq6}
v_{[i]xy}+a_{[i]}v_{[i]x}+b_{[i]}v_{[i]y}+c_{[i]}v_{[i]}=0 
\end{equation}
for $ i\geq1 $ where the coefficients are given by 
\begin{equation} 
\label{s2eq7}
a_{[i]}=a_{[i-1]}-\frac{\partial}{\partial y}\log(h_{[i-1]}), \quad   b_{[i]}=b_{[i-1]},\quad  c_{[i]}=a_{[i]}b_{[i]}+b_{[i]y}-h_{[i-1]}. 
\end{equation}
Here we assume that $ a_{[0]}=a$, $b_{[0]}=b$, $c_{[0]}=c$. Eigenfunctions $ v_{[i]} $ are related by the equations 
\begin{equation} 
\label{s2eq8}
\left(\frac{\partial}{\partial y}+a_{[i]}\right)v_{[i]}=v_{[i+1]}, \quad \left(\frac{\partial}{\partial x}+b_{[i]}\right)v_{[i+1]}=h_{[i]}v_{[i]}. 
\end{equation}
Due to the relation $ c_{[i]}=\frac{\partial}{\partial x}a_{[i]}+a_{[i]}b_{[i]}-h_{[i]} $ system (\ref{s2eq7}) is rewritten as
\begin{equation}
\label{s2eq9} 
a_{[i]}=a_{[i-1]}-\frac{\partial}{\partial y}\log(h_{[i-1]}),\quad h_{[i]}=h_{[i-1]}+a_{[i]x}-b_{[i]y}, \quad b_{[i]}=b.
\end{equation}
Reasonings above define the functions $ a_{[i]}$, $b_{[i]}$, $h_{[i]}$ only for $ i\geq 1 $. However they can be prolonged for $ i\leq 0$ by virtue of the same formulas rewritten as follows
\begin{equation}
\label{s2eq10}
h_{[i-1]}=h_{[i]}-a_{[i]x}-b_y,\quad  a_{[i-1]}=a_{[i]}+\frac{\partial}{\partial y}\log(h_{[i-1]}),\quad b_{[i-1]}=b_{[i]}. 
\end{equation}
Summarising the computations above we get a dynamical system of the form 
\begin{equation}
\label{s2eq11}
\frac{\partial}{\partial y}\log(h_{[i]})=a_{[i]}-a_{[i+1]},\quad a_{[i]x}=h_{[i]}-h_{[i-1]}+b_y,\quad b_{[i]}=b, 
\end{equation} 
which is reduced to the well-known Toda lattice 
\begin{equation}
\label{s2eq12}
\frac{\partial}{\partial y}\log(h_{[i]})=p_{[i]}-p_{[i+1]},\quad p_{[i]x}=h_{[i]}-h_{[i-1]}, 
\end{equation} 
where $ p_{[i]}=a_{[i]}-\tilde{b} $ and $ \tilde{b}_x=b_y. $

Define two linear operators
\begin{equation}
\label{s2eq13}
L_{i}=D_y+a_{[i]}-D_{[i]}, \quad  M_{i}=D_x+b_{[i-1]}-h_{[i-1]}D_{i}^{-1}, 
\end{equation}
where $ D_x, D_y $ are the operators of differentiation with respect to $ x, y $ correspondingly and $ D_{i} $ is the shift operator acting as follows  $ D_{i}a_{[i]}=a_{[i+1]},  D_{i}h_{[i]}=h_{[i+1]} $, etc. We summarize all the reasonings above as a statement.

{\bf Proposition 1}. The operators $ L_{i}, M_{i} $ commute for all $ i $ iff their coefficients satisfy the system (\ref{s2eq9}).

{\bf Corollary}. Equations (\ref{s2eq8}) constitute the Lax pair for the system (\ref{s2eq11}). 

The Laplace $x$-transformation can be interpreted in a similar way.

\section{Laplace cascade for the nonlinear hyperbolic type equations.\\ Formal Lax pairs}

Let us explain how the Laplace cascade is adopted to nonlinear case \cite{Goursat} (see also \cite {SokolovZhiber}). Consider a second order nonlinear hyperbolic type PDE   
\begin{equation}
\label{s3eq1}
u_{xy}=F(x,y,u,u_x,u_y). 
\end{equation}
Its linearization around a solution $ u(x,y) $ derived by substituting $ u=u(x,y,\varepsilon)=u(x,y,0)+\varepsilon v(x,y)+\ldots $ with $\displaystyle{v(x,y)=\frac{\partial u(x,y,\varepsilon)}{\partial\varepsilon}|_{\varepsilon=0}}$ into (\ref{s3eq1}) is an equation of the form 
\begin{equation} 
\label{s3eq2}
v_{xy}+av_x+bv_y+cv=0 
\end{equation}
where the coefficients $ a=-\partial F/\partial u_x$, $b=-\partial F/\partial u_y,$ $c=-\partial F/\partial u $ depend explicitly on the independent variables $ x$, $y $ and the dynamical variables $u$, $u_x$, $u_y$. Let us assign the Laplace sequence (\ref{s2eq8})-(\ref{s2eq11}) to the linearized equation (\ref{s3eq2}). However now instead of the operators $\partial/\partial x$, $\partial/\partial y $ in (\ref{s2eq8})-(\ref{s2eq11}) we use the operators $ D_x$, $D_y $ of the total differentiation with respect to $x$,  and $y$. Denote through $u_i$, $\bar u_i$,  $i=0,1,\ldots $ the $i$-th order derivatives of the variable $ u $ with respect to $ x $ and $ y $ correspondingly 
\begin{equation}
\label{s3eq3} 
D_x^iu=u_i,\quad D_y^iu=\bar u_i. 
\end{equation}
Evidently we have explicit expressions for the operators $ D_x$, $D_y $ acting on the class of smooth functions of $ x, y $ and a finite number of the dynamical variables $ u_i,  \bar u_i $
\begin{equation} 
\label{s3eq4}
D_x= \frac{\partial}{\partial x} + \sum^{\infty}_{i=0}u_{i+1}\frac{\partial}{\partial u_i}+ \sum^{\infty}_{i=1}D_y^{i-1}(F)\frac{\partial}{\partial \bar u_i}, 
\end{equation}
\begin{equation} 
\label{s3eq5}
D_y= \frac{\partial}{\partial y} + \sum^{\infty}_{i=0}\bar u_{i+1}\frac{\partial}{\partial \bar u_i}+ \sum^{\infty}_{i=1}D_x^{i-1}(F)\frac{\partial}{\partial u_i}.
\end{equation}
The Laplace invariants corresponding to the equation (\ref{s3eq2}) are evaluated as 
\begin{equation} 
\label{s3eq6}
h_{[0]}=D_x(a)+ab-c, \quad k_{[0]}=D_y(b)+ab-c.
\end{equation}
The linear system (\ref{s2eq8}) in this case converts into 
\begin{equation} 
\label{s3eq7}
(D_y+a_{[i]})v_{[i]}=v_{[i+1]}, \quad (D_x+b_{[i]})v_{[i+1]}=h_{[i]}v_{[i]}.
\end{equation}
The coefficients $ a_{[i]},  b_{[i]},  h_{[i]} $ are evaluated due to the equations
\begin{equation} 
\label{s3eq8}
\begin{array} {l}
D_y(\log(h_{[i]}))=a_{[i]}-a_{[i+1]}, \quad D_x(a_{[i]})=h_{[i]}-h_{[i-1]}+D_y(b),\\
b_{[i]}=b,\quad a_{[0]}=a.
\end{array}
\end{equation}
All of the mixed derivatives $ u_{xy},  u_{xxy},  \ldots $ are replaced by means of the equation (\ref{s3eq1}) and its differential consequences.

Infinite-dimensional system (\ref{s3eq7}) defines a sequence of the linear operators 
\begin{equation}
\label{s3eq9}
L_{i}=D_y+a_{[i]}-D_{i}, \quad  M_{i}=D_x+b_{[i-1]}-h_{[i-1]}D_{i}^{-1}, 
\end{equation}
satisfying the commutativity conditions 
\begin{equation} 
\label{s3eq10}
\forall i \quad [L_{i},M_{i}]=0.
\end{equation}
Thus one can define a pair of commuting operators $L_{i}$, $M_{i}$ depending on an integer parameter $i$ for an arbitrarily chosen equation (\ref{s3eq1}). Roughly speaking the sequence of the commuting operators (\ref{s3eq9}), (\ref{s3eq10}) recovers equation (\ref{s3eq1}). 
Hence system (\ref{s3eq7}) defines a (formal) Lax pair for the arbitrary (generally non-integrable) equation (\ref{s3eq1}). It is not very surprising since for non-integrable case system (\ref{s3eq7}) is of infinite dimension.  However as it is approved below by several examples for the integrable case the system is either finite (Liouville type equations) or admits a finite dimensional reduction (sine-Gordon type equations). 

{\bf Example 1}. As an illustrative example of the non-integrable equation with the infinite dimensional Lax pair consider the equation 
\begin{equation} 
\label{s3eq11}
u_{xy}=u^2.
\end{equation}
For its linearization 
\begin{equation} 
\label{s3eq12}
v_{xy}=2uv
\end{equation}
we have $a_{[0]}=b_{[0]}=0$, $h_{[0]}=h_{[-1]}=2u$, $a_{[1]}=-\frac{u_y}{u}$, $h_{[1]}=u+\frac{u_xu_y}{u^2}$. One can find all of the coefficients $a_{[j]}$, $h_{[j]}$ due to the equations (\ref{s3eq8}) as functions of the dynamical variables and therefore define completely the system (\ref{s3eq7}). Now go back, suppose that the system evaluated above is consistent and show that its consistency defines uniquely equation (\ref{s3eq11}). Indeed the consistency of (\ref{s3eq7}) implies $D_x(a_{[1]})=h_{[1]}-h_{[0]}$ equivalent to $(-\frac{u_y}{u})_x=\frac{u_xu_y}{u^2}-u$ which gives (\ref{s3eq11}).

{\bf Example 2}. As an  example with the finite system (\ref{s3eq7}) we take the Liouville equation
\begin{equation} 
\label{s3eq13}
u_{xy}=e^{u}
\end{equation}
for which $ h_{[0]}=h_{[-1]}=e^{u} $ and $ h_{[1]}=h_{[-2]}=0. $ For $ i>1 $ and $ i<-2 $ the Laplace invariants $ h_{[i]} $ are not defined. The coefficient $ a_{[i]} $ is defined only for the following three values of $ i: a_{[1]}=-u_y, a_{[0]}=0, a_{[-1]}=u_y. $ Evidently $ b_{[0]}=0. $ System (\ref{s3eq7}) for the equation (\ref{s3eq13}) contains only seven equations
\begin{eqnarray*}
(D_y-u_y)v_{[1]}=v_{[2]}, \quad D_xv_{[2]}=h_{[1]}v_{[1]} \\
D_yv_{[0]}=v_{[1]}, \quad D_xv_{[1]}=h_{[0]}v_{[0]} \\
(D_y+u_y)v_{[-1]}=v_{[0]}, \quad D_xv_{[0]}=h_{[-1]}v_{[-1]} \\
D_xv_{[-1]}=h_{[-2]}v_{[-2]}
\end{eqnarray*}

There is a freedom in choosing of $ v_{[2]}$, $v_{[-2]}.$ Put them equal to zero. Then the obtained system gives the Lax pair for (\ref{s3eq13})
\begin{equation*} 
\Psi_x=A\Psi, \quad  \Psi_y=B\Psi,
\end{equation*}
where $ \Psi=(v_{[1]}, v_{[0]}, v_{[-1]})^{T} $ and

\begin{eqnarray*} 
A=\left( \begin {array}{ccc} 0&e^{u}&0\\ 0&0&e^{u} \\ 0&0&0 \end{array}\right),   
B=\left( \begin {array}{ccc} u_y&0&0\\ 1&0&0 \\ 0&1&-u_y \end{array}\right).
\end{eqnarray*} 

Remark that there are some degenerate cases where the commutativity condition of the operators (\ref{s3eq9}) defines not exactly the initial equation (\ref{s3eq1}) but some other equation connected with (\ref{s3eq1}) by a Miura type transformation. Illustrate it with the following example.

{\bf Example 3}. Consider the equation
\begin{equation} 
\label{s3eq14}
u_{xy}=e^{u+u_x}.
\end{equation}
The coefficients of its linearization 
\begin{equation} 
\label{s3eq15}
v_{xy}=e^{r}(v+v_x)
\end{equation}
depend on $r=u+u_x$. Thus the commutativity condition of the operators (\ref{s3eq9}) assigned to (\ref{s3eq14}) implies the equation $r_{xy}=e^r+r^{r_x}$, connected with (\ref{s3eq14}) by a very simple Miura type transformation $u_x+u=r$.

\subsection{A more symmetrical form of the Laplace sequence}

Let us change the dependent variables in the system (\ref{s3eq7}) to make formulas more symmetrical. Introduce new dependent variables $w_{[i]}$,  $i\in(-\infty,\infty)$ in such a way that $ w_{[i]}=v_{[i]} $ for $ i\geq0 $ and $ w_{[i]}=h_{[-1]}h_{[-2]}...h_{[i]}v_{[i]} $ for $ i\leq-1. $

Then the set of equations (\ref{s3eq7}), (\ref{s3eq8}) is changed to the form below, where $ i\geq0: $
\begin{eqnarray} 
\label{s3.1eq3}
\begin{array}{l}
(D_y+a_{[i]})w_{[i]}=w_{[i+1]}, \quad (D_x+b_{[0]})w_{[i+1]}=h_{[i]}w_{[i]}, \\
D_y(\log(h_{[i]}))=a_{[i]}-a_{[i+1]}, \quad  D_x(a_{[i]})=h_{[i]}-h_{[i-1]}+D_y(b_{[0]}), \\
(D_x+\hat{b}_{[-i]})w_{[-i]}=w_{[-i-1]}, \quad (D_y+a_{[0]})w_{[-i-1]}=h_{[-i-1]}w_{[-i]}, \\
\hat{b}_{[-i-1]}=\hat{b}_{[-i]}-D_x(\log(h_{[-i-1]})), \quad \hat{b}_{[0]}=b_{[0]}
\end{array}
\end{eqnarray}

\subsection{Sine-Gordon equation}

More than two decades ago an important property of the Laplace invariants of the Liouville type integrable equations has been observed  \cite{SokolovZhiber}, \cite{AndersonKamran}. It was proved that the hyperbolic equation (\ref{s3eq1}) is an integrable equation of the Liouville type if and only if the set of its Laplace invariants  is terminated on both sides. Mention also recent results obtained in \cite{Smirnov2015}. The problem of describing the properties of the Laplace invariants characterizing the sine-Gordon type integrable PDE is discussed in  \cite{SokolovZhiber2001}. Our investigation convinces that a connection between the sine-Gordon type equations and the Laplace cascade is clearly formulated in terms of the cascade eigenfunctions. Let us explain our observation with an example.

We consider the sine-Gordon equation
\begin{equation} 
\label{s3.1eq1}
u_{xy}=\sin u.
\end{equation}
It can be shown that the Laplace invariants $ h_{[i]} $ for the linearized equation
\begin{equation} 
\label{s3.1eq2}
v_{xy}=(\cos u) v
\end{equation}
do not vanish identically for any integer $ i. $ Thus system (\ref{s3eq7}) provides an infinite-dimensional Lax pair for (\ref{s3.1eq1}). Below we show that for this case (\ref{s3eq7}) admits a finite-dimensional reduction.
Bring (\ref{s3eq7}) to the symmetric form (\ref{s3.1eq3}).

{\bf Proposition 2}. The system (\ref{s3.1eq3}) corresponding to the sine-Gordon equation (\ref{s3.1eq1}) with $w_{[0]}=v$, $a_{[0]}=0$, $b_{[0]}=0$, $c_{[0]}=-\cos u$ is consistent with the following cutting off boundary conditions
\begin{equation}
\label{s3.1eq4} 
\begin{array}{l}
w_{[2]}=\alpha(-1)w_{[-1]}+\alpha(0)w_{[0]}+\alpha(1)w_{[1]}, \\
w_{[-2]}=\beta(-1)w_{[-1]}+\beta(0)w_{[0]}+\beta(1)w_{[1]}
\end{array}
\end{equation}
where
\begin{equation} 
\label{s3.1eq5}
\begin{array}{l}
\alpha(-1)=-\lambda\frac{u_y}{\sin u}, \quad  \alpha(0)=\lambda, \quad  \alpha(1)=\frac{u_y}{\cos u \sin u}, \\[10pt]
\beta(-1)=\frac{u_x}{\cos u \sin u}, \quad \beta(0)=\lambda^{-1}, \quad \beta(1)=-\lambda^{-1}\frac{u_x}{\sin u},
\end{array}
\end{equation}
$\lambda$ is a complex parameter.

Sketch of proof. Look for the functions (\ref{s3.1eq5}) providing the consistency of the following overdetermined system of equations obtained from (\ref{s3.1eq3}) by imposing (\ref{s3.1eq4})
\begin{equation} 
\label{s3.1eq6}
\left \{
\begin{array}{l}
w_{[1]y}=(\alpha(1)-a_{[1]})w_{[1]}+\alpha(0)w_{[0]}+\alpha(-1)w_{[-1]}, \\
w_{[0]y}=w_{[1]},\\
w_{[-1]y}=h_{[-1]}w_{[0]},
\end{array} 
\right.
\end{equation}
and
\begin{equation}
\label{s3.1eq7}
\left \{
\begin{array}{l}
w_{[1]x}=h_{[0]}w_{[0]}, \\
w_{[0]x}=w_{[-1]},\\
w_{[-1]x}=\beta(1)w_{[1]}+\beta(0)w_{[0]}+(\beta(-1)-\hat{b}_{[-1]})w_{[-1]}.
\end{array} 
\right.
\end{equation}
The compatibility conditions $ (w_{[i]x})_y=(w_{[i]y})_x $ generate a system of nonlinear equations for the functions
$ \alpha(j), \beta(j) $ searched:
\begin{equation}
\label{s3.1eq8} 
\begin{array}{l}
D_y(\beta(-1))+\beta(1)\alpha(-1)=h_{[-2]},\\
D_y(\beta(0))+\beta(1)\alpha(0)+\beta(-1)h_{[-1]}=0,\\
D_y(\beta(1))+\beta(1)(\alpha(1)-a_{[1]})+\beta(0)=0,\\
D_x(\alpha(-1))+\alpha(-1)(\beta(-1)-\hat{b}_{[-1]})+\alpha(0)=0,\\
D_x(\alpha(0))+\alpha(-1)\beta(0)+\alpha(1)h_{[0]}=0,\\
D_x(\alpha(1))+\alpha(-1)\beta(1)=h_{[1]}.
\end{array}
\end{equation}
Here all the given coefficients $ a_{[1]}$, $\hat{b}_{[-1]}$, $h_{[-2]}$, $h_{[-1]}$, $h_{[0]}$, $h_{[1]} $ are linear functions of the derivatives $u_x$, $u_y$:
\begin{equation} 
\label{s3.1eq9}
\begin{array}{l}
h_{[0]}=h_{[-1]}=\cos u, \quad  h_{[1]}=h_{[-2]}=\frac{1}{\cos u}+\frac{u_x u_y}{\cos^2 u},\\
a_{[1]}=u_y \tan u, \quad \hat{b}_{[-1]}=u_x\tan u, 
\end{array}
\end{equation}
therefore we can assume that $ \alpha(j), \beta(j) $ also linearly depend on the first derivatives of $u$
\begin{equation}
\label{s3.1eq10} 
\begin{array}{l}
\alpha(j)=\alpha(j,u,u_y)=p(j,u)u_y+q(j,u), \\
\beta(j)=\beta(j,u,u_x)=r(j,u)u_x+s(j,u), \quad j=1,0,-1.
\end{array}
\end{equation}
Substitute expressions (\ref{s3.1eq9}), (\ref{s3.1eq10}) into equations (\ref{s3.1eq8}) and then compare the coefficients before the independent combinations of the dynamical variables $ u_xu_y, u_x, u_y $. As a result one gets twenty four equations for twelve  functions $p(j,u)$, $q(j,u)$,  $r(j,u)$,  $s(j,u)$, $j=1,0,-1$, depending on $u$ only. We write down explicitly only a part of the equations since the others are obtained from these by applying the replacement $p(j)\leftrightarrow r(-j)$,  $q(j)\leftrightarrow  s(-j)$.

\begin{equation}
\label{s3.1eq11} 
\begin{array}{l}
\frac{dr(-1)}{du}+r(1)p(-1)=\frac{1}{\cos^2 u}, \quad \frac{ds(-1)}{du}+s(1)p(-1)=0, \\
r(1)q(-1)=0, \quad r(-1)\sin u+s(1)q(-1)=\frac{1}{\cos u}, \\
\frac{dr(0)}{du}+r(1)p(0)=0, \quad \frac{ds(0)}{du}+s(1)p(0)=0, \\
r(-1)\cos u+r(1)q(0)=0, \quad r(0)\sin u+s(-1)\cos u+s(1)q(0)=0,\\
\frac{dr(1)}{du}-r(1)\tan u+r(1)p(1)=0, \quad \frac{ds(1)}{du}-s(1)\tan u+s(1)p(1)=0,\\
r(0)+r(1)q(1)=0, \quad r(1)\sin u+s(0)+s(1)q(1)=0.
\end{array}
\end{equation}
By solving the overdetermined system of equations (\ref{s3.1eq11}) we find explicit expressions (\ref{s3.1eq5}) for the coefficients of the constraint (\ref{s3.1eq4}). Now the systems (\ref{s3.1eq6}), (\ref{s3.1eq7}) can be rewritten as follows 

\begin{equation}
\label{s3.1eq12} 
\Psi_x=A\Psi, \quad \Psi_y=B\Psi
\end{equation}
where $ \Psi=(w_{[1]},w_{[0]}, w_{[-1]})^T $ and
\begin{equation}
\label{s3.1eq13} 
A=\left( \begin {array}{ccc} 0&\cos u&0\\ 0&0&1 \\ \frac{-u_x}{\lambda\sin u}&\frac{1}{\lambda}&u_x \cot u \end{array}\right), \,  
B=\left( \begin {array}{ccc} u_y\cot u& \lambda & \frac{-\lambda u_y}{\sin u}\\ 1&0&0 \\ 0& \cos u &0 \end{array}\right).
\end{equation}
It is easily checked that (\ref{s3.1eq12}), (\ref{s3.1eq13}) defines the Lax pair for the sine-Gordon equation (\ref{s3.1eq1}). We failed to reduce it to the well-known usual one found in \cite{Ablowitz73}.

\section{Invariant manifolds of the hyperbolic type PDE}

Let us recall the definition of the invariant manifold of the hyperbolic type equation (\ref{s3eq1}). Consider an equation of the form 
\begin{equation}
\label{s4eq1} 
G(x,y,u_k,u_{k-1}, \ldots u, \bar{u}_1, \bar{u}_2, \ldots \bar{u}_m)=0.
\end{equation}
Note that $ G $ depends on $ x, y $ and a set of the dynamical variables $ u, u_1, \bar{u}_1, \ldots$, where $u_j=\frac{\partial^j u}{\partial x^j}$, $\bar u_j=\frac{\partial^j u}{\partial y^j}$. Take the differential consequences of (\ref{s4eq1})
\begin{equation}
\label{s4eq2} 
G_1(x,y,u_{k+1}, \ldots u, \bar{u}_1, \bar{u}_2, \ldots \bar{u}_m)=0,
\end{equation}
\begin{equation}
\label{s4eq3} 
G_2(x,y,u_k, \ldots u, \bar{u}_1, \bar{u}_2, \ldots \bar{u}_{m+1})=0,
\end{equation}
where $ G_1, G_2 $ are evaluated by applying the operators $ D_x,D_y: G_1=D_xG, G_2=D_yG $ and subsequent replacement of the mixed derivatives by means of the equation (\ref{s3eq1}) and its differential consequences. Equation (\ref{s4eq1}) defines an invariant manifold for (\ref{s3eq1}) if the following equation is satisfied
\begin{equation}
\label{s4eq4} 
D_xD_yG|_{(3.1), (4.1) - (4.3)}=0.
\end{equation}

{\bf Example 4}. Show that equation 
\begin{equation}
\label{s4eq5} 
u_{xx}+\frac{1}{2} u^2_x \tan u=0
\end{equation}
defines an invariant manifold for the sine-Gordon equation (\ref{s3.1eq1}). Here\\ $G=u_{xx}+\frac{1}{2} u^2_x \tan u$, $G_1=D_xG=u_{xxx}+\frac{1}{2} u^3_x $ and $ G_2=D_yG=\frac{u_x}{\cos u}+\frac{u^2_xu_y}{2(\cos u)^2}. $ It is easily verified that
$$D_xD_yG=D_x\left( \frac{u_x}{\cos u}+\frac{u^2_xu_y}{2(\cos u)^2}\right)=0 \quad \mbox{mod} ((4.5),G=0,G_1=0,G_2=0).$$
Therefore equation (\ref{s4eq4}) holds and thus (\ref{s4eq5}) defines an invariant manifold for (\ref{s3.1eq1}).

\subsection{From the Laplace cascade to invariant manifolds}

Show that reduced system (\ref{s3.1eq3}), (\ref{s3.1eq4}) is closely connected with the invariant manifolds of the linearized equation (\ref{s3.1eq2}). Indeed, equations (\ref{s3.1eq3}) imply that $w_{[2]}=(D_y+a_{[1]})(D_y+a_{[0]})w_{[0]}$, $w_{[1]}=(D_y+a_{[0]})w_{[0]}$, $w_{[-1]}=(D_x+b_{[0]})w_{[0]}$, $ w_{[-2]}=(D_x+\hat b_{[-1]})(D_x+b_{[0]})w_{[0]}$. Therefore since $a_{[0]}=b_{[0]}=0$, the boundary conditions (\ref{s3.1eq4}) turn into the equations 
$$ (D_y+a_{[1]})D_y w_{[0]}=\alpha(-1)D_x w_{[0]}+\alpha(0) w_{[0]}+\alpha(1) D_y w_{[0]}, $$
$$ (D_x+\hat b_{[-1]})D_x w_{[0]}=\beta(-1)D_x w_{[0]}+\beta(0) w_{[0]}+\beta(1) D_y w_{[0]}. $$
Simplify the equations obtained by using explicit expressions (\ref{s3.1eq5}), (\ref{s3.1eq9}) and find equations
\begin{equation}
\label{s4eq6} 
\fl L_yw_{[0]}:=(D^2_y-u_y \cot u D_y+\lambda \frac{u_y}{\sin u} D_x-\lambda)w_{[0]}|_{(3.17), w_{[0]xy}=(\cos u)w_{[0]}}=0,
\end{equation}
\begin{equation}
\label{s4eq7} 
\fl L_xw_{[0]}:=(D^2_x-u_x \cot u D_x+\lambda^{-1} \frac{u_x}{\sin u} D_y-\lambda^{-1})w_{[0]}|_{(3.17), w_{[0]xy}=(\cos u)w_{[0]}}=0
\end{equation}
which define the invariant manifold for (\ref{s3.1eq2}) discussed in Introduction (see (\ref{s1.sin:inv1}), (\ref{s1.sin:inv2}) above).

This observation leads to an alternative algorithm to look for the Lax pair. Instead of the cutting off boundary conditions to the lattice (\ref{s3.1eq3}) one searches an invariant manifold for the linearized equation (\ref{s3.1eq2}).

By construction we have
\begin{equation} \label{sin:inv3}
Mw_{[0]}:= (D_xD_y  - \cos u)w_{[0]}.
\end{equation}
Commutators of the operators $L_x$, $L_y$, $M$ satisfy the following relations
\begin{eqnarray}
&&[L_x,L_y]=2A_{xy}(\lambda^{-1}L_y-\lambda L_x),\nonumber\\
&&[M,L_x]=B_{xx}L_y-B_{xy}L_x+(A_{xx}-\lambda^{-1} A_{xy})M,\, \label{commutativity}\\
&&[M,L_y]=B_{yy}L_x-B_{xy}L_y+(A_{yy}-\lambda A_{xy})M \nonumber
\end{eqnarray}
where $ {A=\log \cot \frac{u}{2}}$, $B=\log\sin u$, $A_x=D_x(A)$, $A_y=D_y(A)$, $A_{xx}=D_x^2(A)$ and so on. Consequently any element of the Lie ring generated by the operators $L_x$, $L_y$, $M$ is represented as a linear combination of the same three operators.

Linear equations (\ref{s4eq6}), (\ref{s4eq7}) define a manifold parametrized by $w_{[0]}$, $w_{[0]x}$, $w_{[0]y}$ and the dynamical variables $u$, $u_1$, $\bar u_1, ...\,$. 
By applying $D_x$ to the equations (\ref{s4eq6}), (\ref{s4eq7}) and then simplifying due to the equations (\ref{s3.1eq1}), (\ref{s3.1eq2}) one gets another parametrization of the manifold
\begin{equation}\label{sin:inv4}
w_{[0]xxx} = \frac{u_{xx}}{u_x} w_{[0]xx} + (\lambda^{-1} - u^2_x)w_{[0]x} - \lambda^{-1} \frac{u_{xx}}{u_x}w_{[0]},
\end{equation}
\begin{equation}\label{sin:inv5}
w_{[0]y} = -\lambda \frac{\sin u}{u_x}  w_{[0]xx} + \lambda (\cos u) w_{[0]x} + \frac{\sin u}{u_x} w_{[0]},
\end{equation}
where the parameters $w_{[0]}$, $w_{[0]x}$, $w_{[0]xx}$ are taken as independent ones.
It is shown below that this parametrization is closely connected with the Lax pair for the potential KdV equation being a symmetry of the sine-Gordon equation.

\subsection{Evaluation of the invariant manifolds and the Lax pair for the equation  $u_{xy} = f(u) \sqrt{1+u^2_x}$, $f''=\gamma f$}

In this section we construct a Lax pair to the equation  
\begin{equation}		\label{eq12:main}
u_{xy} = f(u) \sqrt{1+u^2_x}, \quad f'' = \gamma f
\end{equation}
found in \cite{SokolovMeshkov}. It is known that S-integrable equation of the form (\ref{eq12:main}) by an appropriate point transformation  can be reduced either to the case $f(u)=u$ or $f(u)=\sin u$ (see \cite{SokolovMeshkov}).
By analogy with the sine-Gordon equation considered in the previous section we look for the invariant manifold of the form 
\begin{equation} \label{eq12:inv}
v_{yy} + a v_y + b v_x + c v = 0
\end{equation}
for the linearized equation
\begin{equation} \label{eq12:lin}
v_{xy} = f'(u)\sqrt{1+u^2_x}v + \frac{f(u) u_x}{\sqrt{1+u^2_x}} v_x.
\end{equation}
Apply the operator $D_x$ to (\ref{eq12:inv}) and rewrite the result as 
\begin{eqnarray} 
\fl v_{xx} =  -\frac{1}{b}\left( 2  v u_x f(u)f'(u)+ v_xf^2(u)+D_x(b) v_x+c v_x+D_x(c) v+D_x(a) v_y\right) \nonumber\\
-\frac{\bigl(u_y u_x v_x+(v_y +a v)(1+u_x^2)\bigr) f'(u)+\bigl(p u_y v(1+ u_x^2)+a u_x v_x\bigr) f(u)}{b \sqrt{1+u^2_x}}. \label{eq12:eq1}
\end{eqnarray}

Now apply $D_y$ to (\ref{eq12:eq1}), simplify the result due to the equations above and get an equation of the form
\begin{equation} \label{eq12:eq2}
v_{yy} + \tilde{a} v_y + \tilde{b} v_x + \tilde{c} v = 0,
\end{equation}
with the coefficients $\tilde{a}$, $\tilde{b}$, $\tilde{c}$ depending on a finite number of the dynamical variables. According to the definition of the invariant manifold equations (\ref{eq12:inv}) and (\ref{eq12:eq2}) should coincide. This fact implies a system of three equations on the sought functions $a$, $b$, $c$
\begin{eqnarray} 
\fl \Bigl(2 f(u) u_x b f'(u)+D_x(c) b-D_y(b) D_x(a)+D_yD_x(a) b-a b D_x(a)\Bigr)\sqrt{1+u_x^2}\nonumber  \\
 +(2 p  u_y b u_x^2-D_x(a) u_x  b+2 p  u_y b)f(u)-(u_x^2+1)D_y(b) f'(u)  = 0,\label{eq12:eq1:coeff:vy}
 \end{eqnarray}
 \begin{eqnarray}
 \fl \Bigl((-D_y(b) u_x^2+a b-D_y(b)) f(u)^2+u_y b (3+2 u_x^2) f'(u) f(u)\Bigr)\sqrt{1+u_x^2}\nonumber\\
-\bigl( b^2 D_x(a)-D_y(c) b-D_yD_x(b) b+D_y(b) c+D_y(b) D_x(b)\bigr)(1+u_x^2)^{3/2} \nonumber\\
+(u_{xx} b^2+(p u_y^2 u_x b+D_y(a) u_x b-D_y(b) a u_x)(1+u^2_x)) f(u)\nonumber\\
+u_x (1+u_x^2) (b^2 u_x+u_{yy} b+a u_y b-D_y(b) u_y) f'(u)=0,\label{eq12:eq1:coeff:vx}
\end{eqnarray}
\begin{eqnarray}
\fl \Bigl(2 p f(u)^2 u_y u_x b+u_x (a b-2 D_y(b)) f'(u) f(u)+3 f'(u)^2 u_x u_y b\Bigr.\nonumber\\
\Bigl.-D_y(b) D_x(c)-c b D_x(a)+D_yD_x(c) b\Bigr) \sqrt{1+u_x^2}+b (u_x^2+3) f'(u) f(u)^2\nonumber\\
 +\left((-D_y(b) a+D_y(a) b+D_x(b) b+p u_y^2 b)(1+u^2_x)+u_x u_{xx} b^2\right) f'(u)\nonumber\\
+\left(p(  b u_{yy} +a  u_y b+u_x b^2-D_y(b)  u_y)(1+ u_x^2)-D_x(c) u_x b\right) f(u)= 0.\label{eq12:coeff:v}
\end{eqnarray}

Assuming that the searched functions depend only on $u$,$u_x$, $u_y$ i.e.  $a = a(u,u_x,u_y)$, $b=b(u,u_x,u_y)$ and $c = c(u,u_x,u_y)$ substitute these functions into (\ref{eq12:eq1:coeff:vy}), (\ref{eq12:eq1:coeff:vx}) and (\ref{eq12:coeff:v}) and eliminate mixed derivatives of $u$ using (\ref{eq12:main}) from the resulting equations. Thus we obtain three equations of the following form
\begin{equation*}
\alpha_{i}(u,u_x,u_y)u_{xx}u_{yy} + \beta_{i}(u,u_x,u_y)u_{xx} + \gamma_{i}(u,u_x,u_y)u_{yy} + \delta_{i}(u,u_x,u_y) = 0,
\end{equation*}
$i=1,2,3$. These relations are satisfied only if the following conditions:
\begin{equation} \label{eq12:mainsys}
\fl \alpha_{i}(u,u_x,u_y) = 0, \quad \beta_{i}(u,u_x,u_y) = 0, \quad \gamma_{i}(u,u_x,u_y) = 0,\quad \delta_{i}(u,u_x,u_y) = 0
\end{equation}
hold identically for all values $u$, $u_x$ and $u_y$, $i=1,2,3$. Here
\begin{eqnarray*} 
\alpha_{1} & = & (b a_{u_xu_y}-a_{u_x} b_{u_y})\sqrt{1+u^2_x},\\
 \alpha_{2}  &= &( b b_{u_xu_y}-b_{u_x} b_{u_y})(1+u^2_x)^{3/2},  \\
  \alpha_{3}  &= & (b c_{u_xu_y}-c_{u_x} b_{u_y})\sqrt{1+u^2_x}, \\
\beta_{1} &= & \left(b c_{u_x}+b u_y a_{u u_x}-b_{u} u_y a_{u_x}-a b a_{u_x}\right) \sqrt{1+u_x^2}\nonumber\\
& & +f(u) (1+u_x^2) (b a_{u_x u_x}-b_{u_x} a_{u_x}),\\
\beta_{2} &=& -(1+u_x^2) (b_{u} u_y b_{u_x}-b u_y b_{u u_x}+b^2 a_{u_x}) \sqrt{1+u_x^2}\\
&&+f(u) \left((b b_{u_xu_x}-b^2_{u_x})(1+u^2_x)^2 + b b_{u_x}u_x(1+u^2_x)+b^2\right),\\
\beta_{3} &=& (b u_y c_{u u_x}-c b a_{u_x}-b_{u} u_y c_{u_x}) \sqrt{1+u_x^2}\\
&&+(1+u_x^2) (b c_{u_x u_x}-b_{u_x} c_{u_x}) f(u)+b (b_{u_x}(1+u^2_x)+b u_x) f'(u),\\
\gamma_{1} &=& u_x (-a_{u} b_{u_y}+b a_{u u_y}) \sqrt{1+u_x^2}\\
&&+(1+u_x^2) \bigl((b a_{u_y u_y}-b_{u_y} a_{u_y}) f(u)-f'(u) b_{u_y}\bigr),\\
\gamma_{2} &=& (1+u_x^2) \Bigl[(-b_{u_y} b_{u} u_x+b b_{u u_y} u_x-f(u)^2 b_{u_y}-c b_{u_y}+b c_{u_y}) \sqrt{1+u_x^2}\Bigr.\\
&&+\bigl((b b_{u_y u_y}-b_{u_y}^2)(1+u^2_x)-a u_x b_{u_y}+u_x b a_{u_y}\bigr) f(u)\\
&&+u_x (b-b_{u_y} u_y) f'(u)\Bigl.\Bigr],\\
\gamma_{3} &=& u_x (-c_{u} b_{u_y}-2 b_{u_y} f(u) f'(u)+b c_{u u_y}) \sqrt{1+u_x^2}\\
&&+(1+u_x^2) \bigl((p b-b_{u_y} c_{u_y}+b c_{u_y u_y}-p u_y b_{u_y}) f(u)\nonumber\\
&&+(-a b_{u_y}+b a_{u_y}) f'(u)\bigr),\\
\delta_{1} &=& \Bigl[(1+u_x^2) (-b_{u_x} a_{u_y}+b a_{u_x u_y}) f(u)^2+(-u_x^2 b_{u_x}+2 u_x b-b_{u_x}) f'(u) f(u)\Bigr.\nonumber\\
&&\Bigl.-u_x (b_{u} u_y a_{u}+a b a_{u}-b c_{u}-b u_y a_{u u})\Bigr] \sqrt{1+u_x^2}\\
&&+b (u_y a_{u u_y}+ c_{u_y}+2 p u_y -a  a_{u_y}+ a_{u u_x} u_x)(1+u^2_x)f(u)\\
&&\Bigl(b a_{u}-(b_{u_x} a_{u} u_x+b_{u} u_y a_{u_y})(1+u^2_x)\Bigr) f(u)\\
&&+(1+u_x^2) (b a_{u_x} u_x-b_{u} u_y+b a_{u_y} u_y) f'(u),\\
\delta_{2} &=& \Bigl[ (bb_{u_xu_y}-b_{u_x}b_{u_y})(1+u^2_x)^2f^2(u)\\
&& + (1+u^2_x)\bigl(   b a_{u_x}u_x - a b_{u_x}u_x + b b_{u_y}u_x - b_u u_y\bigr)f^2(u)\Bigr. \\
&& + ab f^2(u)+u_y (2 u_x^2 b+3 b-u_x^3 b_{u_x}-u_x b_{u_x}) f'(u) f(u)\\
&&\Bigl.-(1+u_x^2) (-b u_y b_{u u} u_x+b^2 a_{u} u_x+b_{u}^2 u_y u_x+c b_{u} u_y-b c_{u} u_y)\Bigr] \sqrt{1+u_x^2}\\
&&+(1+u_x^2) \Bigl[ (b c_{u_x}+b b_{u}-b^2 a_{u_y} +b  b_{u u_y}u_y -b_{u}  b_{u_y}u_y +b b_{u u_x} u_x \\
&&-c b_{u_x}-b_{u_x} b_{u} u_x -b_{u_x}f^3(u))(1+u^2_x) +u_xu_y(ba_u-ab_u+pbu_y)\\
&&\Bigl.+((b b_{u_y} u_y+b b_{u_x} u_x)(1+u^2_x)+b^2 u_x^2+u_x a u_y b-u_x b_{u} u_y^2) f'(u)\Bigl],\\
\delta_{3} &=& \Bigl[(1+u^2_x)\bigl((b c_{u_x u_y}-p u_y b_{u_x}-b_{u_x} c_{u_y})+2 p u_y u_x b\bigr)f^2(u)\Bigr.\\
&&+\bigl( (b a_{u_x}+b b_{u_y}-a b_{u_x})(1+u^2_x) -2 u_x b_{u} u_y+u_x a b\bigr) f'(u) f(u)\\
&&\Bigl.+3 f'^2(u)b u_x u_y -u_x (c b a_{u}+b_{u}c_{u} u_y -b  c_{u u}u_y)\Bigr] \sqrt{1+u_x^2}\\
&&+(3 b+ b u_x^2-2b_{u_x} u_x^3 -2 b_{u_x} u_x) f'(u) f^2(u)\\
&&+\bigl( (1+u^2_x)(p u_x b^2+b u_y c_{u u_y}-c b a_{u_y}-b_{u} u_y c_{u_y}-b_{u_x} c_{u} u_x-p u_y^2 b_{u} \bigr.\\
&&\bigl.+b c_{u u_x} u_x+a p u_y b) +b c_{u}\bigr)f(u)\\
&&-(1+u_x^2) (a b_{u} u_y-b c_{u_x} u_x-b b_{u} u_x-b a_{u} u_y-b c_{u_y} u_y-p bu_y^2 ) f'(u).
\end{eqnarray*}
Thus the problem is reduced to a system of equations (\ref{eq12:mainsys}). 

We look for the functions $a$, $b$ and $c$  depending on the variable $u_y$ linearly:
\begin{eqnarray*}
a = a_1(u,u_x)u_y + a_2(u,u_x), \\
b = b_1(u,u_x)u_y + b_2(u,u_x), \\
c = c_1(u,u_x)u_y + c_2(u,u_x).
\end{eqnarray*}
Then equation $\alpha_2 =  0$ is essentially simplified $b_2 (b_1)_{u_x} - b_1 (b_2)_{u_x} = 0$. Assume that $b_2 \equiv 0$, then
\begin{equation} \label{eq12:b}
b = b_1(u,u_x)u_y.
\end{equation}
From equations $\alpha_1 = 0$ and $\alpha_3 = 0$ we obtain 
\begin{equation*}
u_y a_{u_x u_x} - a_{u_x} = 0, \quad u_y c_{u_x u_x} - c_{u_x} = 0.
\end{equation*}
Assume that $a_{u_x} = c_{u_x} \equiv 0$ then
\begin{equation} \label{eq12:ac}
a = a_1(u) u_y + a_2(u), \quad c = c_1(u) u_y + c_2(u).
\end{equation}

Substituting the functions (\ref{eq12:b}), (\ref{eq12:ac}) into $\gamma_1 = 0$ (see (\ref{eq12:mainsys} above)) we obtain the following equation
\begin{equation*}
\fl b_1(u,u_x)\left[(f'(u)+a_1(u)f(u))(1+u^2_x)+\sqrt{1+u^2_x}u_xa'_2(u)\right]) = 0.
\end{equation*}
Since functions $f$, $a_1$ and $a_2$ depend only on $u$, we obtain 
\begin{equation*}
f'(u)+a_1(u)f(u)=0, \quad a'_2(u) = 0.
\end{equation*}
Consequently
\begin{equation*}
a_1(u) = -\frac{f'(u)}{f(u)}, \quad a_2(u) = a_3,
\end{equation*}
where $a_3$ is an arbitrary constant.
Then from the equality $\gamma_3 = 0$ we get
\begin{equation*}
c_1(u) = -a_3\frac{f'(u)}{f(u)}, \quad c_2(u) = -f^2(u) + c_3,
\end{equation*}
where $c_3$ is an arbitrary constant. 
 Analyzing the equation $\delta_1 = 0$ we define that 
\begin{equation*}
b_1 = \frac{b_3}{f(u) \sqrt{1+u^2_x}},
\end{equation*}
where $b_3 \neq 0$ is an arbitrary constant.
From the equation $\delta_2 = 0$ we obtain that $a_3 =0$ and $c_3 = -b_3$.
It is easily checked that equalities $\beta_{i}=0$, $i=1,2,3$, $\gamma_2=0$, $\delta_3 = 0$ are automatically satisfied.

Thus summarizing the reasonings above we can claim that equations 
\begin{equation}\label{eq12:vyy1}
v_{yy} - \frac{f'(u)}{f(u)} u_y v_y + \frac{\lambda u_y}{f(u) \sqrt{1+u^2_x}} v_x - (f^2(u) + \lambda) v = 0,
\end{equation}
\begin{equation}
\label{eq12:vxx1}
v_{xx} -\left(\frac{f'(u)}{f(u)}+\frac{u_{xx}}{u_x^2+1}\right) u_x v_x + \frac{u_x \sqrt{u_x^2+1}}{\lambda f(u)} v_y - \left( \frac{u_x^2+1}{\lambda}\right) v = 0
\end{equation}
define an invariant manifold for the linearized equation (\ref{eq12:lin}). Here $\lambda$ is the spectral parameter.

Derive the Lax pair for the equation (\ref{eq12:main}) from the invariant manifold (\ref{eq12:vyy1}), (\ref{eq12:vxx1}). To this end evaluate the Laplace sequence of the form (\ref{s3.1eq3}) for the linearized equation (\ref{eq12:lin}). In what follows we will need in explicit expressions for several first coefficients of the system (\ref{s3.1eq3})
\begin{eqnarray}\label{coeff}
a_{[0]}=a_{[-1]}=-\frac{u_x f(u)}{\sqrt{u_x^2+1}}, \quad b_{[0]}=b_{[1]}=b_{[-1]}=0, \nonumber \\
k_{[0]}=f'(u) \sqrt{u_x^2+1}, \quad h_{[0]}=-\frac{u_xx f(u)}{(u_x^2+1)^\frac{3}{2}}+\frac{f'(u)}{\sqrt{u_x^2+1}}, \\
\fl a_{[1]}=\frac{(u_x^2+1)f(u) (\gamma  u_y \sqrt{u_x^2+1}-f'(u)) + u_x u_{xx} f^2(u)-f'(u) u_y u_{xx} \sqrt{u_x^2+1} }{\sqrt{u_x^2+1} (f'(u)(u_x^2+1) - f(u)u_{xx} )}. \nonumber 
\end{eqnarray}
The constraints (\ref{eq12:vyy1}), (\ref{eq12:vxx1}) generate the cutting off boundary conditions of the form 
\begin{equation}\label{eq12:bc}
\left \{ 
\begin {array}{l} 
v_{[2]}=\alpha(1) v_{[1]}+\alpha(0) v_{[0]}+\alpha(-1) v_{[-1]}, \\ 
v_{[-2]}=\beta(1) v_{[1]}+\beta(0) v_{[0]}+\beta(-1) v_{[-1]}  
\end{array}
\right.
\end{equation}
imposed on the infinite system (\ref{s3.1eq3}). Find the coefficients $\alpha(i)$ and $\beta(i)$ in the relation (\ref{eq12:bc}). Evidently equations (\ref{s3.1eq3}) imply
\begin{equation}\label{eq12:bc-diff}
(D_y+a_{[1]})(D_y+a_{[0]})v_{[0]}=v_{[2]}, \\ 
(D_x+b_{[-1]})(D_x+b_{[0]})v_{[0]}=v_{[-2]}.  
\end{equation}
Set $v:=v_{[0]}$ and simplify (\ref{eq12:bc-diff}) by virtue of (\ref{eq12:bc}) and the relations $v_{[1]}=(D_y+a_{[0]})v$ and $v_{[-1]}=(D_x+b_{[0]})v$. As a result we obtain
\begin{equation}
\label{eq12:vyy2}
\fl \left. 
\begin {array}{l}
v_{yy}+(a_{[0]}+a_{[1]}-\alpha(1))v_y-\alpha(-1)v_x+(a_{[0]y}+a_{[0]}a_{[1]}-a_{[0]}\alpha(1)-\alpha(0)-\alpha(-1)b_{[0]})v=0, \\
v_{xx}+(b_{[-1]}-\beta(-1))v_x-\beta(1)v_y+(b_{[0]x}+b_{[0]}b_{[-1]}-a_{[0]}\beta(1)-\beta(0)-\beta(-1)b_{[0]})v=0, 
\end{array}
\right.
\end{equation} 
The last equations should coincide with (\ref{eq12:vyy1}), (\ref{eq12:vxx1}). Comparison of the corresponding coefficients allows one to derive explicit formulas for the sought functions $\alpha(i)$ and $\beta(i)$
\begin{equation*}
\left \{ 
\begin {array}{l} 
\alpha(1)=a_{[0]}+a_{[1]}+\frac{u_y f'(u)}{u}, \\ 
\alpha(-1)=-\frac{\lambda u_y}{f(u) \sqrt{u_x^2+1}}, \\
\alpha(0)=f(u)^2+\lambda+a_y-a_{[0]}^2-a_{[0]}\frac{u_y f'(u)}{f(u)}+\frac{\lambda u_y b_{[0]}}{f(u) \sqrt{u_x^2+1}},\\
\end{array}
\right.
\end{equation*}
\begin{equation*}
\left \{ 
\begin {array}{l}
\beta(1)=-\frac{u_x \sqrt{u_x^2+1}}{\lambda f(u)},\\
\beta(-1)=b_{[-1]}+\frac{u_x f'(u)}{f(u)} + \frac{u_x u_{xx}}{u_x^2+1}, \\
\beta(0)=b_{[0]x}-a_{[0]}(-\frac{u_x \sqrt{u_x^2+1}}{\lambda f(u)})-(\frac{u_x f'(u)}{f(u)} + \frac{u_x u_{xx}}{u_x^2+1})b_{[0]}+\frac{u_x^2+1}{\lambda}.
\end{array}
\right.
\end{equation*}
Now the infinite system of equations (\ref{s3.1eq3}) is reduced to a pair of the third order systems of ordinary differential equations
\begin{equation*}
 \left \{ 
\begin {array}{l} 
v_{[0]y}=v_{[1]}-a_{[0]}v_{[0]}, \\ 
v_{[1]y}=(\alpha(1) -a_{[1]}) v_{[1]}+\alpha(0) v_{[0]}+\alpha(-1) v_{[-1]}, \\
v_{[-1]y}=k_{[0]}v_{[0]}-a_{[0]}v_{[-1]},\\
\end{array}
\right.
\end{equation*}
\begin{equation*}
 \left \{ 
\begin {array}{l}
v_{[0]x}=v_{[-1]}-b_{[1]}v_{[1]},\\
v_{[1]x}=h_{[0]}v_{[0]}-b_{[0]}v_{[1]},\\
v_{[-1]x}=\beta(1) v_{[1]}+\beta(0) v_{[0]}+(\beta(-1) - b_{[-1]})v_{[-1]}
\end{array}
\right.
\end{equation*}
which can be specified as follows 
\begin{equation}\label{eq12:dy}
\fl \left( \begin {array}{l} v_{[0]} \\ v_{[1]} \\ v_{[-1]} \end{array}\right)_y=
\left( \begin {array}{ccc} \frac{u_x f(u)}{\sqrt{u_x^2+1}} &1&0\\ \lambda & {\frac{u_y f'(u)}{f(u)} - \frac{u_x f(u)}{\sqrt{u_x^2+1}}} & -\frac{u_y \lambda }{ f(u) \sqrt{u_x^2+1}} \\ f'(u)\sqrt{u_x^2+1} &0& \frac{u_x f(u)}{\sqrt{u_x^2+1}} \end{array}\right)
\left( \begin {array}{l} v_{[0]} \\ v_{[1]} \\ v_{[-1]} \end{array}\right),
\end{equation}

\begin{equation}\label{eq12:dx}
\fl \left( \begin {array}{l} v_{[0]} \\ v_{[1]} \\ v_{[-1]} \end{array}\right)_x=\left( \begin {array}{ccc} 0&0&1\\ {\frac{f'(u)}{\sqrt{u_x^2+1}}-\frac{u_{xx}f(u)}{(u_x^2+1)^{\frac{3}{2}}}}&0&0 \\ \frac{1}{\lambda}&-\frac{u_x \sqrt{u_x^2+1}}{\lambda f(u)}& \frac{u_x f'(u)}{f(u)} + \frac{u_x u_{xx}}{u_x^2+1} \end{array}\right)\left( \begin {array}{l} v_{[0]} \\ v_{[1]} \\ v_{[-1]} \end{array}\right).
\end{equation}

Systems (\ref{eq12:dy}), (\ref{eq12:dx}) define the Lax pair for the equation (\ref{eq12:main}).

\section{Searching the Lax pairs for the evolutionary type integrable equations}

Let us consider the evolutionary type integrable equations, for which the Laplace cascade is not defined. Here we use the scheme set out in the Introduction.

\subsection{Korteweg-de Vries equation.}

As an illustrative example we consider the Korteweg-de Vries equation
\begin{equation}		\label{KdV:main}
u_t = u_{xxx} + u u_x.
\end{equation}
Its linearization evidently has the form
\begin{equation} \label{KdV:lin}
v_t = v_{xxx} + u v_x + u_x v.
\end{equation}
Direct computations show that equation (\ref{KdV:lin}) does not admit any invariant manifold of the form $v_{xx}=a(u,u_x,u_{xx})v_{x}+b(u,u_x,u_{xx})v$ fit for arbitrary solution $u(x,t)$ of (\ref{KdV:main}).

Let us look for the invariant manifold of order three 
\begin{equation} \label{KdV:inv}
v_{xxx} = a v_{xx} + b v_x + c v,
\end{equation}
where the coefficients $a$, $b$, $c$ depend on a finite number of the dynamical variables $u$, $u_1$, $u_2, ...\,$.
According to the definition the following condition
\begin{equation}	\label{KdV:compcond}
(v_{xxx})_t = (v_{t})_{xxx}
\end{equation}
should be valid. 

Replacing in (\ref{KdV:compcond}) $v_{xxx}$ and $v_t$ due to (\ref{KdV:inv}) and (\ref{KdV:lin}) respectively and then comparing the coefficients before the independent variables $v_{xx}$, $v_x$ and $v$ we obtain
the following equations
\begin{eqnarray} 
\fl 3 a D_x(b)+6 u_{xx}+D^3_x(a)+3 D_x(a) b+u_x a+u D_x(a)+3 D_x(a)^2\nonumber\\
+3 a D^2_x(a)+3 a^2 D_x(a)+3 D_x(c)-D_t(a)+3 D^2_x(b) = 0,\label{KdV:coeff:vxx}\\
\fl D^3_x(b)+3 D^2_x(c)+3 b D_x(b)+3 D_x(a) D_x(b)+u D_x(b)+4 u_{xxx}+3 a b D_x(a)\nonumber\\
+3 D^2_x(a) b-3 a u_{xx}-D_t(b)+3 c D_x(a)+2 u_x b=0,\label{KdV:coeff:vx}\\
\fl  3 a c D_x(a)+u_{xxxx}+u D_x(c)+D^3_x(c)+3 D^2_x(a) c-a u_{xxx}+3 D_x(b) c\nonumber\\
-D_t(c)+3 u_x c+3 D_x(a) D_x(c)-b u_{xx}=0.\label{KdV:coeff:v}
\end{eqnarray}
It is reasonable to assume that $a = a(u,u_x,u_{xx})$, $b=b(u,u_x,u_{xx})$ and $c = c(u,u_x,u_{xx})$. We substitute these expressions into (\ref{KdV:coeff:vxx}), (\ref{KdV:coeff:vx}) and (\ref{KdV:coeff:v}) and then exclude all the mixed derivatives of $u$ due to the equation (\ref{KdV:main}). As a result we obtain three equations of the following form
\begin{eqnarray*}
\fl \alpha_{i}(u,u_x,u_{xx},u_{xxx})u_{xxxx} &-\beta_{i}(u,u_x,u_{xx}) u^3_{xxx} - \gamma_{i}(u,u_x,u_{xx}) u^2_{xxx}\\
&-\delta_{i}(u,u_x,u_{xx}) u_{xxx}
 - \epsilon_{i}(u,u_x,u_{xx}) = 0, \quad i=1,2,3.
\end{eqnarray*}
Since $u_{xxxx}$, $u^3_{xxx}$, $u^2_{xxx}$, $u_{xxx}$ are independent variables then these equations split down into fifteen equations as follows
\begin{equation} \label{KdV:mainsys}
\begin{array}{c}
\alpha_{i}(u,u_x,u_{xx},u_{xxx}) = 0, \quad \beta_{i}(u,u_x,u_{xx}) = 0, \quad \gamma_{i}(u,u_x,u_{xx}) = 0,\\
 \delta_{i}(u,u_x,u_{xx}) = 0, \quad \epsilon_{i}(u,u_x,u_{xx})=0
 \end{array}
\end{equation}
hold for all values of $u$, $u_x$ and $u_{xx}$, $i=1,2,3$. Thus the searched coefficients $a$, $b$, $c$ satisfy a highly overdetermined system of differential equations (\ref{KdV:mainsys}).
Specify and analyse the system. It can be verified that the three equations $\beta_i = 0$, $i=1,2,3$ immediately imply $a_{u_{xx} u_{xx} u_{xx}} = 0$, $\beta_{u_{xx} u_{xx} u_{xx}} = 0$ and $c_{u_{xx}u_{xx}u_{xx}} = 0$ therefore
\begin{eqnarray*}
a &= a_1(u,u_x) u^2_{xx}  + a_2(u,u_x)u_{xx} + a_3(u,u_x),\\
b &= b_1(u,u_x) u^2_{xx}  + b_2(u,u_x)u_{xx} + b_3(u,u_x),\\
c &= c_1(u,u_x) u^2_{xx} + c_2(u,u_x)u_{xx} + c_3(u,u_x).
\end{eqnarray*}
Equations $\alpha_i = 0$, $i=1,2,3$ are of the form
 \begin{eqnarray}
a_{u_{xx} u_{xx}} u_{xxx} +a a_{u_{xx}} +  a_{u u_{xx}} u_x + b_{u_{xx}} + a_{u_x u_{xx}}u_{xx} &= 0,
\label{KdV:uxxxx:1}\\
b_{u_{xx} u_{xx}} u_{xxx} + c_{u_{xx}} + b_{u u_{xx}} u_x + b_{u_x u_{xx}}u_{xx} + b a_{u_{xx}} &= 0  ,	\label{KdV:uxxxx:2}\\
3 c_{u_{xx} u_{xx}} u_{xxx}  + 3 c_{u u_{xx}}u_x +  3 c_{u_x u_{xx}}u_{xx} + 3 c a_{u_{xx}} + 1 & = 0 .	\label{KdV:uxxxx:3}
\end{eqnarray}
Then since functions $a$, $b$ and $c$ depend only on the variables $u$, $u_x$ and $u_{xx}$ the coefficients at $u_{xxx}$ vanish, i.e. we get
\begin{eqnarray*}
a = a_1(u,u_x)u_{xx} + a_2(u,u_x), \\
 b = b_1(u,u_x)u_{xx} + b_2(u,u_x),   \\
  c = c_1(u,u_x)u_{xx} + c_2(u,u_x).
\end{eqnarray*}
Substituting $a$, $b$ and $c$ into the equations (\ref{KdV:uxxxx:1}), (\ref{KdV:uxxxx:2}) and (\ref{KdV:uxxxx:3}) we obtain
\begin{eqnarray*}
\bigl(a^2_1+ (a_1)_{u_x}\bigr)u_{xx} + a_1 a_2 + u_x (a_1)_u + b_1 = 0,\\
\bigl( (b_1)_{u_x} + a_1 b_1  \bigr)u_{xx} + c_1 + u_x (b_1)_u + a_1 b_2 = 0,\\
3 \bigl( (c_1)_{u_x} + a_1 c_1  \bigr)u_{xx} + 3 u_{x}(c_1)_u + 3 a_1 c_2 + 1 = 0.
\end{eqnarray*}
Since functions $a_{i}$, $b_{i}$ and $c_{i}$, $i=1,2$ depend only on $u$ and $u_x$  the coefficients at $u_{xx}$ vanish and we get the following system of equations
\begin{equation} 
\eqalign{a^2_1+ (a_1)_{u_x} = 0, \\
 (b_1)_{u_x} + a_1 b_1  = 0, \\
 (c_1)_{u_x} + a_1 c_1 = 0,\\
a_1 a_2 + u_x (a_1)_u + b_1 = 0, \\
c_1 + u_x (b_1)_u + a_1 b_2 = 0 = 0, \\
 3 u_{x}(c_1)_u + 3 a_1 c_2 + 1 = 0.}\label{KdV:eq13}
\end{equation}
Concentrate on the last system. It is easy to check that $a_1 \neq 0$. Assume that $b_1 \neq 0$ and $c_1 \neq 0$. From system (\ref{KdV:eq13}) we find
\begin{equation}
\begin{array}{lll}
\displaystyle  a_1 = \frac{1}{u_x + a_4(u)}, &\displaystyle b_1 = \frac{b_4(u)}{u_x + a_4(u)}, & \displaystyle c_1 = \frac{c_4(u)}{u_x + a_4(u)},\\
\displaystyle a_2 = -\frac{u_x (a_1)_u + b_1}{a_1}, & \displaystyle b_2 = -\frac{c_1 + u_x (b_1)_u}{a_1}, &\displaystyle c_2 = -\frac{1+3u_x (c_1)_u}{3 a_1}.
\end{array}
\end{equation}
Now equations $\gamma_{i} = 0$, $i=1,2,3$ are satisfied automatically. From equation $\delta_1 = 0$ we obtain that $b_4 = b_5$  and $a_4 = a_5 u + a_6$, where $b_5$, $a_5$ and $a_6$ are arbitrary constants. From equation $\delta_2 = 0$ we get $c_4 = \frac{2}{3} u + c_5$, where $c_5$ is an arbitrary constant. Equation $\delta_3 = 0$ implies $a_5 = b_5$. Then from equation $\epsilon_1 = 0$ we obtain $a_5=a_6=b_5 = 0$. It is easy to check that equations $\epsilon_2 = 0$ and $\epsilon_3 = 0$ are identically satisfied.

Thus equation  (\ref{KdV:inv}) is of the form
\begin{equation}\label{KdV:L}
v_{xxx} = \frac{u_{xx}}{u_x} v_{xx} - \left( \frac{2}{3} u + \lambda\right) v_x + \left( \left( \frac{2}{3} u + \lambda \right)\frac{u_{xx}}{u_x} - u_x \right)v.
\end{equation}
Here $\lambda=c_5$ is an arbitrary parameter.

{\bf Proposition 3}. {\it Pair of equations (\ref{KdV:lin}), (\ref{KdV:L}) defines the Lax pair for the KdV equation.}

\subsection{Potential and modified KdV equations}

Concentrate on the potential KdV equation
\begin{equation}		\label{symsin:main}
u_t = u_{xxx} + \frac{1}{2} u^3_x.
\end{equation}
Its linearization 
\begin{equation} \label{symsin:lin}
v_t = v_{xxx} + \frac{3}{2}w^2 v_x,\, \mbox{where}\, w=u_x
\end{equation}
does not admit any second order invariant manifold of the necessary form $v_{xx}=a(u,u_x,u_{xx},...)v_x+b(u,u_x,u_{xx},...)v$. However it admits a third order invariant manifold given by
\begin{equation} \label{symsin:inv1}
v_{xxx} = \frac{w_{x}}{w} v_{xx} - \left( w^2 + \lambda\right) v_x + \lambda \frac{w_{x}}{w}v.
\end{equation}
Here $\lambda$ is an arbitrary parameter.
The consistency condition of the equations (\ref{symsin:lin}), (\ref{symsin:inv1}) is equivalent to the mKdV equation
\begin{equation}		\label{mKdV}
w_t = w_{xxx} + \frac{1}{2}w^2 w_x
\end{equation}
connected with (\ref{symsin:main}) by a very simple substitution $w=u_x$. In other words (\ref{symsin:lin}), (\ref{symsin:inv1}) define the Lax pair for the equation (\ref{mKdV}) as well.

Now let us return to the sine-Gordon equation (\ref{s3.1eq1}). Recall that equation (\ref{symsin:main}) is a symmetry of the sine-Gordon equation. 	It is easily seen that equation (\ref{symsin:inv1}) coincides with (\ref{sin:inv4}) up to the notations.

\subsection{Lax pairs for the KdV type equations from Svinolupov-Sokolov list}

Consider the following two third order differential equations 
\begin{eqnarray}
u_t = u_{yyy} - \frac{\gamma}{2} u^3_y - \frac{3}{2} f^2(u) u_y, \label{eq12:sym1}\\
u_{\tau} = u_{xxx} - \frac{3u_x u^2_{xx}}{2(1+u^2_{x})} - \frac{\gamma}{2} u^3_x \label{eq12:sym2}
\end{eqnarray}
possessing infinite hierarchies of conservation laws  \cite{SvinolupovSokolov}. As it is established in \cite{SokolovMeshkov} these equations are symmetries of the equation (\ref{eq12:main}). We have proved above in \S 4 that equations 
\begin{equation}	\label{eq12:inv:y}
v_{yy} - \frac{f'(u)}{f(u)} u_y v_y + \frac{\lambda u_y}{f(u) \sqrt{1+u^2_x}} v_x - (f^2(u) + \lambda) v = 0,
\end{equation}
\begin{equation}	\label{eq12:inv:x}
v_{xx} -\left(\frac{f'(u)}{f(u)}+\frac{u_xx}{u_x^2+1}\right) u_x v_x + \frac{u_x \sqrt{u_x^2+1}}{\lambda f(u)} v_y - \left( \frac{u_x^2+1}{\lambda}\right) v = 0
\end{equation}
define an invariant manifold for the linearized equation (\ref{eq12:lin}). It is reasonable to expect that invariant manifolds for the linearizations 
\begin{equation}  \label{eq12:sym1:lin}
v_t = v_{yyy} - \frac{3}{2} (\gamma u^2_y+ f^2(u))v_y - 3 f(u) f'(u) u_y v
\end{equation}
and
\begin{equation} \label{eq12:sym2:lin}
v_{\tau} = v_{xxx} - \frac{3 u_x u_{xx}}{1+u^2_x} v_{xx} - \frac{3}{2} \left( \frac{(1-u^2_x)u^2_{xx}}{(1+u^2_x)^2} + \gamma u^2_x \right) v_x
\end{equation}
of the symmetries (\ref{eq12:sym1})\ and (\ref{eq12:sym2}) are closely connected with the same manifold.
Indeed by applying the operators $D_y$ and $D_x$ to the equations (\ref{eq12:inv:y}) and respectively to (\ref{eq12:inv:x}) one can deduce the following two third order ordinary differential equations
\begin{eqnarray}  
\fl v_{yyy} - \frac{u_{yy}}{u_{y}} v_{yy} - \bigl( \gamma u^2_y + f^2(u) + \lambda  \bigr) v_y \nonumber\\
+\left( \frac{(f^2(u)+\lambda)u_{yy}}{u_y} - 3 f(u)f'(u)u_y  \right) v = 0,   \label{eq12:vyyy}
\end{eqnarray}
\begin{eqnarray}
\fl v_{xxx} - \frac{(1+3 u^2_x)u_{xx}}{(1+u^2_x)u_x} v_{xx}  \nonumber\\
-\left( \lambda^{-1} + \gamma u^2_x + \frac{u_x u_{xxx}}{1+u^2_x} - \frac{3 u^2_x u^2_{xx}}{(1+u^2_x)^2} \right) v_x +\frac{u_{xx}}{u_x} \lambda^{-1} v=0.  \label{eq12:vxxx}
\end{eqnarray}
It is easily checked by a direct computation that equations (\ref{eq12:vyyy}) and (\ref{eq12:vxxx}) define invariant manifolds for (\ref{eq12:sym1:lin}) and (\ref{eq12:sym2:lin}) correspondingly.
Conclude the reasonings with the following statement.

{\bf Proposition 4}. {\it 1) Linear equations (\ref{eq12:sym1:lin}), (\ref{eq12:vyyy}) define the Lax pair for the equation (\ref{eq12:sym1});

\noindent
2) linear equations (\ref{eq12:sym2:lin}), (\ref{eq12:vxxx}) define the Lax pair for the equation (\ref{eq12:sym2}).}


\subsection{Volterra type integrable chains}

In this section we discuss the semi-discrete equations of the form $\frac{\partial}{\partial t}u_n=f(u_{n+1},u_n,u_{n-1})$ with the sought function $u=u_n(t)$, depending on the discrete $n$ and continuous $t$.
The direct method for constructing the Lax pairs through linearization can be applied to the discrete models  as well. As illustrative examples we consider the modified Volterra chain
\begin{equation} \label{sd2}
\frac{d p_{n}}{dt} = -p^2_{n}(p_{n+1}-p_{n-1})
\end{equation}
and the equation 
\begin{equation} \label{sd0}
\frac{d u_{n}}{dt} = \frac{1}{u_{n+1}-u_{n-1}}
\end{equation}
found in \cite{Yamilov83}. 
These two equations are related to each other by a very simple Miura type transformation
\begin{equation}  \label{sd1}
\frac{d u_{n}}{dt} = p_{n}, \qquad p_{n}=\frac{1}{u_{n+1}-u_{n-1}}.
\end{equation}

Note that the coefficients of the linearization 
\begin{equation}	\label{sd3}
\frac{d v_{n}}{dt} = -p^2_{n} (v_{n+1}-v_{n-1})
\end{equation}
of the equation (\ref{sd0}) depend on the variable $p_n$. This explains why we study these two equations together. Look for the invariant manifold of the third order: 
\begin{equation}	\label{sd4}
v_{n+2} = a v_{n+1} + b v_{n} + c v_{n-1}
\end{equation}
to the equation (\ref{sd3}) with the coefficients $a$, $b$, $c$ depending on a finite set of the dynamical variables $p_n,p_{n\pm 1},...\,$. Actually we suppose that (\ref{sd4}) defines an invariant manifold for any choice of the solution   $p=p_n(t)$ to the equation (\ref{sd2}). 

The coefficients $a$, $b$, $c$ are found from the equation
\begin{equation}  \label{sd5}
\frac{d}{dt} ( a v_{n+1} + b v_{n} + c v_{n-1}) = D_{n}^2 \left( -p^2_{n} (v_{n+1}-v_{n-1}) \right).
\end{equation}
Studying the equation (\ref{sd5}) we assume that the variables $\left\{p_k\right\}_{k=-\infty}^{\infty}$, $v_n$, $v_{n+1}$, $v_{n-1}$ are independent dynamical variables. Omitting the simple but tediously long computations we give only the answer:
$$a=-\frac{p_{n}}{p_{n+1}} + \frac{\lambda}{p^2_{n+1}},\quad b=1 - \frac{\lambda}{p_{n}p_{n+1}},\quad c=\frac{p_{n}}{p_{n+1}}.$$
Therefore the invariant manifold searched is of the form:
\begin{eqnarray}
\fl v_{n+2} = \left( -\frac{p_{n}}{p_{n+1}} \right. & + \left.\frac{\lambda}{p^2_{n+1}}\right) v_{n+1}\nonumber\\
&+ \left( 1 - \frac{\lambda}{p_{n}p_{n+1}} \right) v_{n} 
+ \frac{p_{n}}{p_{n+1}} v_{n-1}. \label{invMv}
\end{eqnarray}

{\bf Proposition 5}. {\it The consistency condition of the equations (\ref{sd3}) and (\ref{invMv}) coincides with the equation (\ref{sd2}).}

\section{Construction of the recursion operators and conservation laws  via newly found Lax pairs}

It was observed that the Lax pairs for the integrable equations found above essentially differ from their classical counterparts. In this section we discuss some useful properties of the newly found Lax pairs. We show, for instance, that they provide a very convenient tool for searching the recursion operators and conservation laws for integrable models. As illustrative examples we take the KdV and pKdV equations, the Volterra type chain (\ref{sd0}) etc. We show that the equation of the invariant manifold to the linearized equation is easily transformed into the recursion operator.

\subsection{Evaluation of the recursion operators for the KdV type equations}

Let us start with the KdV equation (\ref{s1kdv}). We can rewrite equation (\ref{s1KdV:L}) of the invariant manifold in the following form
\begin{eqnarray}
\left( D^3_x - \frac{u_{xx}}{u_x} D^2_x + \frac{2 u}{3} D_x + u_x - \frac{2 u u_{xx}}{3 u_x} \right) v = \lambda u_x D_x \frac{1}{u_x} v. \label{6.1}
\end{eqnarray}
We now multiply (\ref{6.1}) from the left by the operator $ u_x D^{-1}_x \frac{1}{u_x} $ and obtain a formal eigenvalue problem of the form 
\begin{eqnarray}
R v = \lambda v \label{6.2}
\end{eqnarray} 
for the operator
\begin{eqnarray}
R =  u_x D^{-1}_x \frac{1}{u_x} \left( D^3_x - \frac{u_{xx}}{u_x} D^2_x + \frac{2 u}{3} D_x + u_x - \frac{2 u u_{xx}}{3 u_x}\right)   \label{6.3}
\end{eqnarray}
An amazing fact is that $ R $ coincides with the recursion operator for the KdV equation. Indeed, we have 
\begin{eqnarray*}
R v =  u_x D^{-1}_x \left( \frac{v_{xxx} u_x - u_{xx} v_{xx}}{u^2_x} + \frac{2}{3} \frac{u v_x}{u_x} + v - \frac{2}{3} \frac{u u_{xx} v}{u^2_x} \right) = \\ v_{xx} + u_x D^{-1}_x \left( \frac{2}{3} \frac{u v_x}{u_x} + v - \frac{2}{3} \frac{u u_{xx} v}{u^2_x} \right). 
\end{eqnarray*}
It can be simplified due to the relation 
\begin{eqnarray*}
\frac{u v_x}{u_x} -  \frac{u u_{xx} v}{u^2_x} =\left( \frac{u v}{u_x}\right)_{x} - v
\end{eqnarray*}
and reduced to the form
\begin{eqnarray*}
R v = \left( D^2_x + \frac{2}{3} u + \frac{1}{3} u_x D^{-1}_x\right) v.
\end{eqnarray*}
In a similar way we derive from (\ref{symsin:inv1}) the recursion operator to the potential KdV equation (\ref{symsin:main}). Indeed rewrite (\ref{symsin:inv1}) as follows 
\begin{eqnarray}
\left( D^3_x - \frac{u_{xx}}{u_x} D^2_x + u^2_x D_x \right) v = - \lambda u_x D_x \frac{1}{u_x} v.   \label{6.4}
\end{eqnarray}
Equation (\ref{6.4}) implies 
\begin{eqnarray*}
R v = - \lambda v
\end{eqnarray*}
where 
\begin{eqnarray*}
R = u_x D^{-1}_x \frac{1}{u_x} \left( D^3_x - \frac{u_{xx}}{u_x} D^2_x + u^2_x D_x \right).
\end{eqnarray*}
Simplify the expression for $Rv$:
\begin{eqnarray*}
R v= u_x D^{-1}_x \left( \frac{v_{xxx} u_x - u_{xx} v_{xx}}{u^{2}_x} + u_x v_x \right) = v_{xx} + u_x D^{-1}_x u_x D_x v.
\end{eqnarray*}
Apparently operator  $ R = D^2_x + u_x D^{-1}_x u_x D_x $ coincides with the recursion operator for (\ref{symsin:main}).

Invariant manifolds (\ref{eq12:vyyy}) and (\ref{eq12:vxxx}) for the linearized equations (\ref{eq12:sym1:lin}) and (\ref{eq12:sym2:lin}) allow constracting of the recursion operators 
\begin{eqnarray*}
R = D^2_y-\gamma u^2_y - \frac{f^2(u)}{u_y} + u_y D^{-1}_y \left( \gamma u_{yy} - f f'\right)
\end{eqnarray*}
and
\begin{eqnarray*}
R = D^2_x - \frac{2 u_x u_{xx}}{1 + u^2_x} + u_x D^{-1}_x \left( \frac{u_{xxx}}{1 + u^2_x} + \frac{u_x u^2_{xx}}{(1 + u^2_x)^2} -  \gamma u_x \right) D_x
\end{eqnarray*}
for the Svinolupov-Sokolov equations (\ref{eq12:sym1}) and (\ref{eq12:sym2}) respectively.

\subsection{Recursion operators via the Lax pair for a Volterra type chain}

We proceed with an example of the discrete model (\ref{sd0}). Let us write equation (\ref{invMv}) defining the invariant manifold for (\ref{sd3}) as follows 
\begin{eqnarray}
\left( D^2_n + \frac{p_n}{p_{n+1}} D_n - 1 -\frac{p_n}{p_{n+1}} D^{-1}_n \right) v_n = \frac{\lambda}{p_{n+1}} \left( D_n - 1 \right) \frac{v_n}{p_n}, \label{6.2.1}
\end{eqnarray}
where $ D_n $ is the shift operator acting due to the rule $ D_n a(n) = a(n+1) $. We multiply (\ref{6.2.1}) from the left by the factor $ p_n \left(D_n - 1\right)^{-1} p_{n+1} $ and then get 
\begin{eqnarray}
R v = \lambda v \label{6.2.2}
\end{eqnarray}
where $ R = p_n \left(D_n - 1\right)^{-1} p_{n+1} \left( D^2_n + \frac{p_n}{p_{n+1}} D_n - 1 -\frac{p_n}{p_{n+1}} D^{-1}_n \right). $
After some elementary transformations the operator $ R $ is reduced to the form 
\begin{eqnarray}
R = p^2_n \left(D_n +D^{-1}_n\right) + 2 p_n p_{n-1} + 2 p_n \left(D_n - 1\right)^{-1} \left(p_{n-1} - p_{n+1}\right) \label{6.2.3}
\end{eqnarray}
Operator $ R $ given by (\ref{6.2.3}) defines the recursion operator for the chain (\ref{sd0}). For instance, by applying the operator $ R $ to $ \frac{1}{u_{n+1} - u_{n-1}} $ we obtain the r.h.s. of the equation constructed in \cite{Tongas}
\begin{eqnarray}
\frac{\partial}{\partial\tau}u_{n} = \frac{1}{\left(u_{n+1} - u_{n-1}\right)^2} \left(\frac{1}{u_{n+2} - u_{n}} + \frac{1}{u_{n} - u_{n-2}}\right) \label{6.2.4}
\end{eqnarray}
as a fifth-point symmetry of the equation (\ref{sd0}). Note that the whole hierarchy of the symmetries for the chain (\ref{sd0}) is described in \cite{svinin}.

\subsection{Conservation laws via the Lax pair for a Volterra type chain}

Let us consider an equation of the form 
\begin{equation} 
\frac{d u_{n}}{dt} = \frac{1}{u_{n+1}-u_{n-1}}.  \label{maindiscrEq}
\end{equation}
Due to the relations (\ref{sd1}) between $u$ and $p$ a pair of the equations (\ref{sd3}), (\ref{invMv}) define a Lax pair to the chain (\ref{maindiscrEq}) as well. 

Rewrite the scalar Lax pair (\ref{sd3}), (\ref{invMv}) in the matrix form
\begin{equation} \label{VLax}
y_{n+1} = f y_n, \qquad \frac{{\rm d} y_{n}}{\rm{dt}} = g y_n,
\end{equation}
where
\begin{eqnarray}  \label{VLgf}
f = \left( \begin{array}{ccc}
-\frac{p_n}{p_{n+1}} + \frac{\lambda}{p^2_{n+1}} & 1- \frac{\lambda}{p_n p_{n+1}} & \frac{p_n}{p_{n+1}}\\
1 & 0 & 0\\
0 & 1 & 0
\end{array} \right),\\
g= \left( \begin{array}{ccc}
p_n p_{n+1} - \lambda & \frac{p_{n+1}}{p_n} \lambda & -p_n p_{n+1}\\
-p^2_n & 0 & p^2_n\\
p_{n-1} p_n & -\frac{p_{n-1}}{p_n} \lambda & -p_{n-1}p_n + \lambda
\end{array} \right).
\end{eqnarray}
To construct the conservation laws, we apply the method of the formal diagonalization suggested in  \cite{HabLOMI, HabYang} and developed in \cite{HabPoptsova}.

The first equation in (\ref{VLax}) has singular point $\lambda = \infty$ ($f$ has a pole at $\lambda = \infty$). It can be checked that the potential $f$ is represented as follows 
\begin{equation*}
f = \alpha Z \beta,
\end{equation*}
where 
\begin{eqnarray}
 \alpha = \left(\begin{array}{ccc}
\frac{1}{p^2_{n+1}} - \frac{p_n}{p_{n+1}}\lambda^{-1} & 0 & 0\\
\lambda^{-1} & \frac{p_{n+1}}{p_n} & 0\\
0 & 1 & 1
\end{array} \right), \\
 \beta = \left(  \begin{array}{ccc}
 1 & - \frac{p_{n+1}}{p_n} & \frac{p_n p_{n+1}}{\lambda - p_n p_{n+1}}\\
 0 & 1 & -\frac{p^2_n}{\lambda - p_n p_{n+1}}\\
 0 & 0 & \frac{p^2_n \lambda}{\lambda - p_n p_{n+1}}
 \end{array} \right)
\end{eqnarray}
are analytic and non-degenerate around $\lambda = \infty$, $Z$ is a diagonal matrix of the form
\begin{equation}
Z = \left( \begin{array}{ccc}
\lambda & 0 & 0\\
0 & 1 & 0\\
0 & 0 & \lambda^{-1}
\end{array} \right).
\end{equation}
The change of variables $\psi = \beta y$ is reducing the first system in (\ref{VLax}) to the special form
\begin{equation}  \label{Vsf}
\psi_{n+1} = PZ \psi_n,
\end{equation}
where
\begin{eqnarray*}
\fl P = D_n(\beta) \alpha = 
 \left(  \begin{array}{ccc}
\frac{1}{p^2_{n+1}} - \frac{p_n + p_{n+2}}{p_{n+1}} \lambda^{-1} & -\frac{p_{n+2}}{p_n} \left( 1 - \frac{p_n p_{n+1}}{\lambda - p_{n+1} p_{n+2}}\right) & \frac{p_{n+1} p_{n+2}}{\lambda - p_{n+1} p_{n+2}}\\
\lambda^{-1} & \frac{p_{n+1}}{p_n} \left( 1 - \frac{p_n p_{n+1}}{\lambda - p_{n+1} p_{n+2}} \right) & -\frac{p^2_{n+1}}{\lambda - p_{n+1} p_{n+2}}\\
0 & \frac{p^2_{n+1} \lambda}{\lambda - p_{n+1} p_{n+2}} & \frac{p^2_{n+1} \lambda}{\lambda - p_{n+1} p_{n+2}}
\end{array} \right).
\end{eqnarray*}
The function $P(\lambda)$ together with $P^{-1}(\lambda)$ are analytic around
$\lambda = \infty$:
\begin{equation}  \label{VP}
P(\lambda) = \sum_{i \geq 0}^{\infty} P^{(i)} \lambda^{-i},
\end{equation}
where
\begin{eqnarray*}
\fl P^{(0)} = \left(  \begin{array}{ccc}
\frac{1}{p^2_{n+1}} & -\frac{p_{n+2}}{p_n} & 0\\
0 & \frac{p_{n+1}}{p_n} & 0\\
0 & p^2_{n+1} & p^2_{n+1}
\end{array} \right), \quad
P^{(1)} = \left(  \begin{array}{ccc}
-\frac{p_{n+1}(p_n + p_{n+2})}{p^2_{n+1}} & p_{n+1} p_{n+2} & p_{n+1} p_{n+2}\\
1 & -p^2_{n+1} & - p^2_{n+1}\\
0 & p^3_{n+1}p_{n+2} & p^3_{n+1}p_{n+2}
\end{array}
\right),\\
\fl P^{(2)} = \left( \begin{array}{ccc}
0 & p^2_{n+1} p^2_{n+2} & p^2_{n+1} p^2_{n+2} \\
0 & -p^3_{n+1} p_{n+2} & -p^3_{n+1} p_{n+2}\\
0 & p^4_{n+1}p^2_{n+2} & p^4_{n+1}p^2_{n+2}
\end{array} \right), \quad
P^{(3)} = \left( \begin{array}{ccc}
0 & p^3_{n+1} p^3_{n+2} & p^3_{n+1} p^3_{n+2} \\
0 & -p^4_{n+1} p^2_{n+2} & -p^4_{n+1} p^2_{n+2}\\
0 & p^5_{n+1}p^3_{n+2} & p^5_{n+1}p^3_{n+2}
\end{array} \right) \quad \mbox{etc.}
\end{eqnarray*}
The leading principal minors of $P(\lambda)$ 
\begin{eqnarray*}
\det_1 P (\lambda = \infty) = \frac{1}{p^2_{n+1}}, \qquad \det_2 P (\lambda = \infty) = \frac{1}{p_n p_{n+1}} ,\\
\det_3 P(\lambda = \infty) = \det P(\lambda = \infty) = \frac{p_{n+1}}{p_n} 
\end{eqnarray*}
do not vanish if the variable $p_n$ satisfy the inequality $p_n \neq 0$ for all $n$.  According to the Proposition 1 in \cite{HabYang} the first system in (\ref{VLax}) can be diagonalized, i.e. there exist formal series
\begin{eqnarray} 
T = T^{(0)} + T^{(1)} \lambda^{-1} + T^{(2)} \lambda^{-2} + \cdots, \label{VT}\\
h = h^{(0)} + h^{(1)} \lambda^{-1} + h^{(2)} \lambda^{-2} + \cdots  \label{Vh}
\end{eqnarray}
such that the formal change of the variables $\psi = T \varphi $ converts the system (\ref{Vsf}) to the system of the diagonal form
\begin{equation} \label{diagformV}
\varphi_{n+1} = h Z \varphi_n.
\end{equation}
Thus we see that the formal change of variables $y=R \varphi = \beta^{-1} T \varphi $, reduces the first system in investigated Lax pair (\ref{VLax}) to the form (\ref{diagformV}). By construction $R = \beta^{-1} T$ is a formal series of the form
\begin{equation} \label{VR}
R = R^{(0)} + R^{(1)} \lambda^{-1} + R^{(2)} \lambda^{-2} + \cdots
\end{equation}
It follows from (\ref{Vh})--(\ref{VR}) that diagonalizable system (\ref{diagformV}) admits an asymptotic representation of the solution to the direct scattering problem
\begin{equation}  \label{Vformalsolution}
y_n(\lambda) = R(n,\lambda)e^{\sum_{s=n_0}^{n-1} \log h(s, \lambda)} Z^n
\end{equation}
with the ``amplitude'' $A = R(n,\lambda)$ and ``phase'' $\phi = n \log Z + \sum_{s=n_0}^{n-1} \log h(s,\lambda)$.

Turn back to the problem of diagonalization. By solving the following equation 
\begin{equation}  \label{VThPeq}
D_n(T) h = P (\lambda) \bar{T}, \qquad \bar{T} = ZTZ^{-1}
\end{equation}
we find the formal series $T$ and $h$
\begin{eqnarray*}
\fl T = \left(\begin{array}{ccc}
1 & 0 & 0\\
0 & 1 & 0\\
0 & p_{n-1} p_ n & 	1
\end{array} \right) +
\left( \begin{array}{ccc}
0 & \frac{p^2_{n+1} p_{n+2}}{p_{n}} & 0\\
p^2_{n} & 0 & 0\\
0 & p_n p^2_{n-1}(p_n - p_{n-2}) & 0
\end{array}
\right) \lambda^{-1} \\
+ \left( \begin{array}{ccc}
0 & \frac{p^2_{n+1} p^2_{n+2} (p_{n+1}+p_{n+3})}{p_n} & 0\\
p^3_n (2 p_{n-1} + p_{n+1}) & 0 & p_n p_{n+1}\\
p^4_n p^2_{n-1} & T^{(2)}_{32} & 0
\end{array} \right) \lambda^{-2} + \cdots,
\end{eqnarray*}
\begin{eqnarray*}
\fl h = \left( \begin{array}{ccc}
\frac{1}{p^2_{n+1}} & 0 & 0\\
0 & \frac{p_{n+1}}{p_n} & 0\\
0 & 0 & p^2_{n+1}
\end{array} \right)
 + \left(  \begin{array}{ccc}
 -\frac{p_{n+1} (p_n + p_{n+2})}{p^2_{n+1}} & 0 & 0\\
 0 & \frac{p^2_{n+1} ( p_{n+2} - p_n)}{p_n} & 0\\
 0 & 0 & p^3_{n+1} (p_n + p_{n+2}) 
 \end{array} \right) \lambda^{-1}  \\
 + \left(  \begin{array}{ccc}
-p_n p_{n+2} & 0 & 0\\
0 & h^{(2)}_{22} & 0\\
0 & 0 & p^4_{n+1}(3 p_n p_{n+2} + p^2_n + p^2_{n+2})
\end{array} \right)  \lambda^{-2} + \cdots
\end{eqnarray*}
where
\begin{eqnarray*}
\fl T^{(2)}_{32} = p_n p_{n-1} (p^2_{n-2} p^2_{n-1} + p^2_{n-2}p_{n-1}p_{n-3} + p^2_{n-1}p^2_n + 2 p^2_{n-1} p_n p_{n-2} - p_{n+2} p^2_{n+1} p_n),\\
h^{(2)}_{22} = -\frac{p^2_{n+1} (-p^2_{n+2} p_{n+3} - p^2_{n+2}p_{n+1} +p^2_n p_{n-1} + p_n p_{n+1} p_{n+2})}{p_n}.
 \end{eqnarray*}
According to the general scheme the second system of the Lax pair (\ref{VLax}) is diagonalized
by the same linear change of the variables $y = R \varphi$, where
\begin{eqnarray}
\fl R= \beta^{-1} T = \left(  \begin{array}{ccc}
1 & -\frac{p_{n+1}}{p_n} & 0\\
0 & 1 & 0\\
0 & -p_{n-1} p_n & p^2_n
\end{array}  \right)  \\
\fl + \left( \begin{array}{ccc}
0 & - \frac{p^2_{n+1} p_{n+2}}{p_n} & p_n p_{n+1}\\
-p^2_n  & p_n p_{n+1} & -p^2_n\\
p_{n-1} p^3_n & -p_{n-1} p^2_n p_{n+1} - p^2_{n-1} p^2_n - p_{n-2} p^2_{n-1} p_n & p^3_n(p_{n+1} + p_{n-1})
\end{array} \right) \lambda^{-1} \\
+ \left( \begin{array}{ccc}
R^{(2)}_{11} & R^{(2)}_{12} & R^{(2)}_{13}\\
R^{(2)}_{21} & R^{(2)}_{22} & R^{(2)}_{23}\\
R^{(2)}_{31} & R^{(2)}_{32} & R^{(2)}_{33}
\end{array} \right) \lambda^{-2} + \cdots,
\end{eqnarray}
\begin{eqnarray*}
R^{(2)}_{11} = p_n p^2_{n+1} p_{n+2}, \\
R^{(2)}_{12} = -\frac{p^2_{n+1} p_{n+2} (p_{n+2} p_{n+3} + p_{n+1} p_{n+2} + p_n p_{n+1})}{p_n},\\
R^{(2)}_{13} = p_n p^2_{n+1} (p_n + p_{n+2}),\\
R^{(2)}_{21} = -p^3_n (2 p_{n-1} + p_{n+1}),\\
R^{(2)}_{22} = p_n p^2_{n+1} p_{n+2} + 3 p_{n-1} p^2_n p_{n+1} + p^2_n p^2_{n+1},\\
R^{(2)}_{23} = -3 p^3_n p_{n+1},\\
R^{(2)}_{31} = 2 p^2_{n-1} p^4_{n} + p_{n-1} p^4_n p_{n+1} + p_{n-2} p^2_{n-1} p^3_{n},\\
R^{(2)}_{32} = -p_{n-1} p_n (3 p_{n-1} p^2_n p_{n+1} + p^2_n p^2_{n+1} + p_{n-2} p_{n-1} p_n p_{n+1} + p^2_{n-2} p^2_{n-1} \\
+ p_{n-3} p^2_{n-2} p_{n-1} + p^2_{n-1} p^2_n + 2 p_{n-2} p^2_{n-1} p_n).
\end{eqnarray*}
This change of the variables reduces the second system in (\ref{VLax}) to the form $\frac{{\rm d} y_n}{{\rm  d} t} = S y_n$ with
\begin{eqnarray*}
\fl  S = -R^{-1} R_t + R^{-1} g R 
 = \left( \begin{array}{ccc}
1 & 0 & 0\\
0 & 0 & 0\\
0 & 0 & -1
\end{array} \right) \lambda\\
+
\left( \begin{array}{ccc}
2 p_n p_{n+1} & 0 & 0\\
0 & p_n (p_{n-1} - p_{n+1}) & 0\\
0 & 0 & -2p_n p_{n+1}
\end{array} \right)  
+\left(\begin{array}{ccc}
S^{(2)}_{11} & 0 & 0\\
0 & S^{(2)}_{22} & 0\\
0 & 0 & S^{(2)}_{33}
\end{array}  \right) \lambda^{-1} + \cdots,
\end{eqnarray*}
where
\begin{eqnarray*}
S^{(2)}_{11} = p_n p_{n+1} (p_{n+1} p_{n+2} + p_{n-1} p_n),\\
S^{(2)}_{22} =  p_n p_{n+1} (p_{n-1} p_n - p_{n+1} p_{n+2}),\\
S^{(2)}_{33} = -p_n p_{n+1} (p_{n+1} p_{n+2} + p_{n-1} p_n).
\end{eqnarray*}
According the paper \cite{HabYang} the equation
\begin{equation*}
D_t \ln h = (D_n - 1) S
\end{equation*}
generates the infinite series of conservation laws for the equations (\ref{sd2}). We write down in an explicit form three conservation laws from the infinite sequence
obtained by the diagonalization procedure
\begin{eqnarray*}
\fl D_t \left(\ln \frac{1}{p_{n+1}}\right) = (D_n - 1) p_n p_{n+1},\\
\fl D_t (-p_{n+1} (p_n + p_{n+2})) = (D_n - 1) p_n p_{n+1} (p_{n-1}p_n + p_{n+1} p_{n+2} ),\\
\fl D_t \left( -2 p_n p^2_{n+1} p_{n+2} - \frac{1}{2} p^2_n p^2_{n+1} - \frac{1}{2} p^2_{n+1} p^2_{n+2} \right) \\
= (D_n - 1) p^2_n p^2_{n+1} (2 p_{n-1}  p_{n+2} + p_{n+1} p_{n+2} + p_{n-1} p_n ).
\end{eqnarray*}
Being rewritten in terms of $u_n$ these relations give the conservation laws for the equation (\ref{maindiscrEq}):
\begin{eqnarray*}
\fl D_t \ln (u_{n+2} - u_n) = (D_n-1)\frac{1}{(u_{n+1}-u_{n-1})(u_{n+2} - u_n)},\\
\fl D_t \frac{u_{n-1}-u_{n+3}}{(u_{n+1}-u_{n-1})(u_{n+2} - u_n)(u_{n+3} - u_{n+1})} \\
= (D_n-1) \left( \frac{1}{(u_n - u_{n-2})(u_{n+1}-u_{n-1})^2 (u_{n+2}-u_n)} \right.\\
\left.+ \frac{1}{(u_{n+1}-u_{n-1})(u_{n+2}-u_n)^2(u_{n+3} - u_{n+1})}  \right),\\
\fl D_t \left( -\frac{2}{(u_{n+1}-u_{n-1})(u_{n+2}-u_n)^2(u_{n+3}-u_{n+1})} \right.\\
 -\frac{1}{2(u_{n+1}-u_{n-1})^2(u_{n+2} - u_n)^2}\\
\left.- \frac{1}{2 (u_{n+2}-u_n)^2 (u_{n+3}-u_{n+1})^2}\right) \\
= (D_n-1) \left( \frac{2}{(u_n - u_{n-2})(u_{n+1}-u_{n-1})^2 (u_{n+2} - u_n)^2 (u_{n+3} - u_{n+1})} \right. \\
+ \frac{1}{(u_{n+1} - u_{n-1})^2 (u_{n+2} - u_n)^3 (u_{n+3} - u_{n+1})} \\
\left.+ \frac{1}{(u_n - u_{n-2})(u_{n+1} - u_{n-1})^3 (u_{n+2} - u_n)^2} \right).
\end{eqnarray*}
These conservation laws coincide with those found earlier (see, for instance, \cite{Yamilov82}, \cite{Yamilov83}).

\section*{Conclusions}

There is a large set of classification methods allowing classes of integrable nonlinear PDEs
and their discrete analogues to be described. For studying the analytical properties of these
equations one needs the Lax pairs. Therefore the problem of creating convenient algorithms
for constructing the Lax pairs is relevant. In the present article such a method is suggested.
For the evolution type integrable equation the Lax pair consists of the linearized equation and
the equation of its invariant manifold. In the case of the hyperbolic equations to obtain the
Lax pair we use the Laplace cascade in addition to the invariant manifold. The method is
applied to equations (\ref{eq12:main}), (\ref{eq12:sym1}) and (\ref{eq12:sym2}) known to be integrable for which the Lax pairs
have not been constructed before.

An interesting observation is connected with the Laplace cascade of the sine-Gordon
equation. It is proved that in this case the cascade admits a finite-dimensional reduction which
generates the Lax pair to the sine-Gordon model. We conjecture that the Laplace cascade
corresponding to any hyperbolic type integrable equation admits a finite-dimensional
reduction.

We considered examples showing that our method leads to true Lax pairs having useful
applications. For the Lax pair of the Volterra type chain we found an asymptotic
eigenfunction which allowed an infinite set of conservation laws to be constructed. It is also
shown that these Lax pairs allow the recursion operators to be constructed describing
infinite hierarchies of the higher symmetries and invariant manifolds for the given nonlinear
equation.
\section*{Acknowledgments}

The authors gratefully acknowledge financial support from a Russian Science Foundation grant (project 15-11-20007).

\section{Appendix. How to look for the invariant manifolds for the linearization of a nonlinear hyperbolic type equation}

It is well known that the linear hyperbolic type equation 
$v_{xy}=p(x,y)v_x+q(x,y)v_y+r(x,y)v$
might admit linear invariant manifolds (see \cite{Kaptsov}). Here we consider equation (\ref{s3eq2})
obtained by linearizing an essentially nonlinear equation (\ref{s3eq1}). We request that at least one of the coefficients $a$, $b$, $c$ in (\ref{s3eq2}) depends on at least one of the dynamical variables $u$, $u_1$, $\bar u_1$. We look for a linear invariant manifold of the form
\begin{equation}\label{s4eq77}
(p(j)D_y^j+p(j-1)D_y^{j-1}+ ... +p(1)D_y+p(0)+q(1)D_x +...+q(k)D_x^{k})v=0.
\end{equation}
It is assumed that all the coefficients $p(i)$ and $q(m)$ can depend on $x$, $y$ and on a finite set of the dynamical variables $u,u_1,\bar u_1,u_{2}, \bar u_{2},...$. More precisely we study a family of equations  (\ref{s3eq2}), (\ref{s4eq77}) depending on a functional parameter $u(x,y)$ such that when $u=u(x,y)$ ranges a set of all solutions to the equation (\ref{s3eq1}) then (\ref{s4eq77}) ranges a set of invariant manifolds for the corresponding equation (\ref{s3eq2}). In this case equation (\ref{s4eq77}) generates an overdetermined system of differential equations for defining the searched coefficients $p(i)$ and $q(m)$. 

As an illustrative example we consider equation (\ref{s3.1eq2}) obtained by linearizing the sine-Gordon equation. We search an invariant manifold for (\ref{s3.1eq2}) in the following form
\begin{equation} \label{sin:inv}
v_{yy} + a v_y + b v_x + c v = 0.
\end{equation}
Here $a$, $b$ and $c$ are functions depending on a finite number of the dynamical variables $u,u_1,\bar u_1,u_{2}, \bar u_{2},...$. 
Apply the operator $D_x$ to (\ref{sin:inv}) and rewrite the obtained result as follows
\begin{eqnarray} \label{sin:eq5}
\fl v_{xx} =
 \frac{1}{b} \left( \sin(u) u_y v-\cos(u) (v_y+a  v)-D_x(a) v_y-D_x(b) v_x-D_x(c) v-c v_x \right).
\end{eqnarray}

Now apply the operator $D_y$ to this equation and simplify the result by means of the equation (\ref{s3.1eq2}). We arrive at the equation
\begin{equation*} 
v_{yy} + \tilde{a} v_y + \tilde{b} v_x + \tilde{c} v = 0,
\end{equation*}
which should coincide according to the definition above with equation (\ref{sin:inv}), i.e. 
\begin{equation*}
 \tilde{a} = a, \quad \tilde{b} = b, \quad \tilde{c} = c.
\end{equation*}
These conditions give rise to the following equalities
\begin{eqnarray} 
\fl D_yD_x(a) b-D_y(b) \cos(u)-2 \sin(u) u_y b+D_x(c) b- D_x(a)(a b+D_y(b)) = 0,\label{sin:coeff:vy}\\
\fl D_yD_x(b) b-D_y(b) D_x(b)+D_y(c) b-D_y(b) c-b^2 D_x(a) = 0,\label{sin:coeff:vx}\\
\fl D_yD_x(c) b-D_y(b) D_x(c)-c b D_x(a)
+ \cos(u)\left(- u_y^2 b+D_y(a)  b+D_x(b)  b-D_y(b) a \right)\nonumber\\
+\sin(u)\left(- u_{yy} b-b^2  u_x-a  u_y b+D_y(b) u_y\right) = 0.\label{sin:coeff:v}
\end{eqnarray}

Assume that $a = a(u,u_x,u_y)$, $b=b(u,u_x,u_y)$ and $c = c(u,u_x,u_y)$ and substitute these functions into (\ref{sin:coeff:vy}), (\ref{sin:coeff:vx}) and (\ref{sin:coeff:v}). Eliminating the mixed derivatives of $u$ due to equation (\ref{s3.1eq1}) we obtain three equations of the following form
\begin{equation*}
\alpha_{i}(u,u_x,u_y)u_{xx}u_{yy} + \beta_{i}(u,u_x,u_y)u_{xx} + \gamma_{i}(u,u_x,u_y)u_{yy} + \delta_{i}(u,u_x,u_y) = 0,
\end{equation*}
$i=1,2,3$. Since the functions $\alpha_{i}$, $\beta_{i}$, $\gamma_{i}$ and $\delta_{i}$, $i=1,2,3$ depend only on $u$, $u_x$ and $u_y$, we should have the coefficients at $u_{xx} u_{yy}$, $u_{xx}$, $u_{yy}$ and remaining terms equal to zero, i.e. 
\begin{eqnarray} 
\fl \alpha_{i}(u,u_x,u_y) &= 0, \qquad \beta_{i}(u,u_x,u_y) = 0,  \qquad
\gamma_{i}(u,u_x,u_y) &= 0,\qquad \delta_{i}(u,u_x,u_y) = 0\label{sin:mainsys}
\end{eqnarray}
for all $u$, $u_x$ and $u_y$, $i=1,2,3$. Here
\begin{eqnarray} 
\alpha_1 & = & b a_{u_xu_y}-b_{u_y} a_{u_x},\label{sin:alpha1}\\
\alpha_2  &=&  b b_{u_xu_y}-b_{u_y} b_{u_x}, \label{sin:alpha2}\\
 \alpha_3 & =&  b c_{u_xu_y}-c_{u_x} b_{u_y},\label{sin:alpha3} \\
\beta_1 &= & b c_{u_x}-a b a_{u_x}-b_u u_y a_{u_x}-b_{u_x} \sin(u) a_{u_x}\nonumber\\
&&+b a_{u_x u_x} \sin(u)+b a_{u u_x} u_y,\label{sin:beta1}\\
\beta_2 &= &   b b_{u u_x} u_y+b b_{u_x u_x} \sin(u)-b_u u_y b_{u_x}-b^2 a_{u_x}-b_{u_x}^2 \sin(u),\label{sin:beta2}\\
\beta_3 & = & b c_{u_x u_x} \sin(u)+b c_{u u_x} u_y+b b_{u_x} \cos(u)-c b a_{u_x}\nonumber\\
&& -b_{u_x} \sin(u) c_{u_x}-b_u u_y c_{u_x},\label{sin:beta3}\\
\gamma_1 &=&  -\cos(u) b_{u_y}+b \sin(u) a_{u_y u_y}+b u_x a_{u u_y}-b_{u_y} a_u u_x-b_{u_y} a_{u_y} \sin(u),\nonumber\\
\gamma_2 &=& b u_x b_{u u_y}-b_{u_y} b_u u_x+b \sin(u) b_{u_y u_y}-b_{u_y}^2 \sin(u)+b c_{u_y}-c b_{u_y},\nonumber\\
\gamma_3 &=& -b_{u_y} c_u u_x-b \sin(u)-b_{u_y} c_{u_y} \sin(u)+b a_{u_y} \cos(u)+\sin(u) u_y b_{u_y}\nonumber\\
&&+b u_x c_{u u_y}-a \cos(u) b_{u_y}+b \sin(u) c_{u_y u_y},\nonumber\\
\delta_1 &= &(b_{u_x} a_{u_y}-b a_{u_xu_y}) \cos^2(u)- b_{u_x}\cos(u) \sin(u)\nonumber\\
& &+(b a_{u_y} u_y+b a_{u_x} u_x-b_{u} u_y) \cos(u)\nonumber\\
& &+(b a_{u}-2 u_y b+b a_{uu_y} u_y-b_{u} u_y a_{u_y})\sin(u)\nonumber\\
& & + (-a b a_{u_y}-b_{u_x} a_{u} u_x+b u_x a_{uu_x}+b c_{u_y}) \sin(u)\nonumber\\
& &+b u_x a_{uu} u_y-b_{u} u_y a_{u} u_x-a b a_{u} u_x+b a_{u_xu_y}-b_{u_x} a_{u_y}+b c_{u} u_x,\nonumber\\
\delta_2&=& (b_{u_x} b_{u_y}-b b_{u_xu_y}) \cos^2(u)+b (b_{u_x} u_x+b_{u_y} u_y) \cos(u)\nonumber\\
&&+(-b_{u} u_y b_{u_y}-b_{u_x} b_{u} u_x+b u_x b_{uu_x})\sin(u)\nonumber\\
&&+(b b_{uu_y} u_y-b^2 a_{u_y}-c b_{u_x}+b b_{u}+b c_{u_x}) \sin(u)\nonumber\\
&&+b b_{u_xu_y}+b c_{u} u_y-b_{u}^2 u_y u_x-c b_{u} u_y-b^2 a_{u} u_x+b u_x b_{uu} u_y-b_{u_x} b_{u_y},\nonumber\\
   \delta_3 &=& (b_{u_x} c_{u_y}-b c_{u_xu_y}-u_y b_{u_x}) \cos^2(u)+(b a_{u_x}-a b_{u_x}+b b_{u_y}) \cos(u) \sin(u)\nonumber\\
   && +(b b_{u} u_x-a b_{u} u_y+b a_{u} u_y+b c_{u_x} u_x+b c_{u_y} u_y-u_y^2 b) \cos(u)\nonumber\\ 
&& +(b u_x c_{uu_x}-a u_y b+b c_{uu_y} u_y-b_{u_x} c_{u} u_x)\sin(u)\nonumber\\
&&+(-c b a_{u_y}-b_{u} u_y c_{u_y}-b^2 u_x+b c_{u}+u_y^2 b_{u}) \sin(u)\nonumber\\
 &&+b u_x c_{uu} u_y-b_{u} u_y c_{u} u_x-c b a_{u} u_x+u_y b_{u_x}-b_{u_x} c_{u_y}+b c_{u_xu_y}.\nonumber
\end{eqnarray}
Hence the problem of searching the equation (\ref{sin:inv}) is reduced to the system of equations (\ref{sin:mainsys}). 

We now look for the functions $a$, $b$ and $c$  depending linearly on the variable $u_y$:
\numparts
\begin{eqnarray} 
a = a_1(u,u_x)u_y + a_2(u,u_x),\label{sin:abc:a}\\
b = b_1(u,u_x)u_y + b_2(u,u_x), \label{sin:abc:b}\\
c = c_1(u,u_x)u_y + c_2(u,u_x).\label{sin:abc:c}
\end{eqnarray}
\endnumparts
Then equation $\alpha_2 =  0$ (see (\ref{sin:alpha2})) takes the form $b_2 (b_1)_{u_x} - b_1 (b_2)_{u_x} = 0$. This equation is satisfied only if at least one of the following conditions holds
\begin{equation*}
b_1 = 0, \qquad b_2 = 0, \qquad \frac{(b_1)_{u_x}}{b_1} = \frac{(b_2)_{u_x}}{b_2}. 
\end{equation*}
Let us consider the case $b_2 = 0$. Then function $b$ defined by the formula (\ref{sin:abc:b}) takes the form
\begin{equation} \label{sin:b}
b = b_1(u,u_x)u_y.
\end{equation}
Furthermore equations $\alpha_1 = 0$ and $\alpha_3 = 0$ become
\begin{equation*}
u_y a_{u_x u_x} - a_{u_x} = 0, \quad u_y c_{u_x u_x} - c_{u_x} = 0.
\end{equation*}
These equations hold if at least one of the following four conditions is satisfied:
\begin{eqnarray*}
1) a_{u_x} = 0,\, c_{u_x}=0;& &\\
2) a_{u_x} = 0,\, c_{u_x} \neq 0,& \quad & \frac{c_{u_x u_x}}{c_{u_x}} = \frac{1}{u_y};\\
3) c_{u_x} = 0, \, a_{u_x} \neq 0,&  &\frac{a_{u_x u_x}}{a_{u_x}} = \frac{1}{u_y};\\
4) a_{u_x} c_{u_x} \neq 0,&  &\frac{a_{u_x u_x}}{a_{u_x}} = \frac{c_{u_x u_x}}{c_{u_x}}.
\end{eqnarray*}

It can be proved that cases 2)-4) lead to contradiction. Concentrate on the case 1) which implies 
\begin{equation} \label{sin:ac}
a = a_1(u) u_y + a_2(u), \quad c = c_1(u) u_y + c_2(u).
\end{equation}
By substituting the functions (\ref{sin:b}), (\ref{sin:ac}) into equalities $\beta_i = 0$, $i=1,2,3$ (see (\ref{sin:beta1}),(\ref{sin:beta2}) and (\ref{sin:beta3})) one can verify that equation $\beta_1 = 0$ is satisfied.  The equations $\beta_2 = 0$ and $\beta_3 = 0$ lead to
\begin{equation*}
b_1 (b_1)_{u u_x} - (b_1)_u (b_1)_{u_x} = 0 \quad \mbox{and}\quad(\cos u) u^2_y b_1 (b_1)_{u_x} = 0
\end{equation*}
correspondingly. Thus $b_1(u,u_x) = F_1(u)$ for some function $F_1(u)$. The equality $\gamma_1 = 0$ becomes
\begin{equation*}
F_1(u)(\cos u+u_x a'_2(u)+a_1(u)\sin u)=0.
\end{equation*}
Clearly, the functions $a_1$ and $a_2$ are defined by the formulas
\begin{equation*}
a_1(u) = -\mathrm{\cot}u, \quad a_2(u) = C_1,
\end{equation*}
where $C_1$ is an arbitrary constant. According to that we rewrite the equality $\gamma_3 = 0$ as
\begin{equation*}
F1(u)(u_xc'_2(u)+c_1(u)\sin u+\cos uC_1) = 0.
\end{equation*}
Since $u$ and $u_x$ are regarded as independent variables the last equation leads to
\begin{equation*}
c_1(u) =-C_1\mathrm{\cot}u, \quad c_2(u) \equiv C_2,
\end{equation*}
where $C_2$ is an arbitrary constant. Let us turn to the equation $\delta_1 = 0$ which gives:
\begin{equation*}
u^3_y u_x(F_1(u)\cos u+F'_1(u)\sin u) = 0.
\end{equation*}
Integration of the equation leads to
\begin{equation*}
F_1 = \frac{C_3}{\sin u}.
\end{equation*}
Now the equation $\delta_2 = 0$ takes the form
\begin{equation*}
C_3u^2_y\bigl(C_1 u_y \sin u+(C_3 +C_2)\cos u\bigr) = 0
\end{equation*}
We get $C_2 + C_3 = 0$, $C_1 = 0$.  It is easy to check that equalities $\delta_3 = 0$ and $\gamma_2 = 0$ are identically satisfied. Thus we have
\begin{equation*}
a = -u_y \mathrm{\cot}u, \quad b = \lambda\frac{u_y}{\sin u}, \quad c =  - \lambda.
\end{equation*}
Therefore the invariant manifold (\ref{sin:inv}) is of the form
$$v_{yy}-u_y\cot u \,v_y+\lambda\frac{u_y}{\sin u} v_x-\lambda v=0$$
and coincides with (\ref{s4eq6}) while  (\ref{sin:eq5}) gives (\ref{s4eq7}).

\end{document}